\documentclass[preprint,prd,nofootinbib,tightenlines,amsmath]{revtex4}
\usepackage{graphicx}% Include figure files
\usepackage{bm}% bold math
\usepackage{color}
\usepackage{multirow}

\begin{document}
\baselineskip=15pt \parskip=3pt

\vspace*{3em}

\title{Constraining Nonstandard Neutrino-Electron Interactions\\ due to a New Light Spin-1 Boson}

\author{Cheng-Wei Chiang,$^{a,b,c}$, Gaber Faisel,$^{a,d,e}$ Yi-Fan Lin,$^a$ and Jusak Tandean$^{a,e}$}
\affiliation{$^a$Department of Physics and Center for Mathematics and Theoretical Physics,
National Central University, Chungli 320, Taiwan
\medskip \\
$^b$Institute of Physics, Academia Sinica,\\ Taipei 115, Taiwan
\medskip \\
$^c$Physics Division, National Center for Theoretical Sciences,\\ Hsinchu 300, Taiwan
\medskip \\
$^d$Egyptian Center for Theoretical Physics,
Modern University for Information and Technology,\\ Cairo, Egypt
\medskip \\
$^e$Department of Physics and Center for Theoretical  Sciences, National Taiwan
University, Taipei 106, Taiwan \\
$\vphantom{\bigg|_{\bigg|}^|}$}

%\date{\today}

\begin{abstract}
We consider nonstandard interactions of neutrinos with electrons arising from a new light
spin-1 particle with mass of tens of GeV or lower and couplings to the neutrinos and electron.
This boson is not necessarily a gauge boson and is assumed to have no mixing with standard-model
gauge bosons.
Adopting a~model-independent approach, we study constraints on the flavor-conserving and
-violating couplings of the boson with the leptons from a number of experimental data.
Specifically, we take into account the (anti)neutrino-electron scattering and
\,$e^+e^-\to\nu\bar\nu\gamma$\, measurements and keep explicitly the dependence on the new
particle mass in all calculations.
We find that one of the two sets of data can provide the stronger constraints,
depending on the mass and width of the boson.
Also, we evaluate complementary constraints on its separate flavor-conserving couplings to
the electron and neutrinos from other latest experimental results.
\end{abstract}

\maketitle

%%%%%%%%%%%%%%%%%%%%%%%%%%%%%%%%%%%%%%%%%%%%%%%%%%
\section{Introduction\label{intro}}
%%%%%%%%%%%%%%%%%%%%%%%%%%%%%%%%%%%%%%%%%%%%%%%%%%

A growing amount of experimental data has now confirmed that neutrinos possess mass and mix
among themselves~\cite{pdg}.
The masslessness of the neutrinos in the minimal standard model (SM) implies that extra
ingredients beyond them are necessary to account for this observation.
Despite the accumulating knowledge of neutrino properties, the nature of the mechanism
responsible for generating neutrino masses and mixing is still a mystery~\cite{pdg}.
It is generally expected, however, that the underlying new physics would also modify
the structure of the electroweak neutral and charged currents in the SM.
Such modifications in the neutrino sector give rise to the so-called nonstandard
interactions (NSI) of the neutrinos~\cite{Berezhiani:2001rs,Barranco:2007ej,Forero:2011zz}.
In most studies on such NSI, they arise from the exchange of new particles that
are usually assumed to be heavier than the electroweak scale and, thus, lead to effective
four-fermion interactions for low-energy phenomenology.
Nevertheless, it is also feasible that the exchanged new particle is not heavy, {\it e.g.},
in the GeV or sub-GeV regime.
One of the simplest possibilities along this line is that the new particle is a spin-1 boson.

Scenarios beyond the SM involving new spin-1 particles with relatively low masses
have been considered to some extent in various contexts in the literature.
Generally speaking, their existence is not just still compatible with current data,
but also highly desirable, as they may offer explanations for some of the recent
experimental anomalies and unexpected observations.
For instance, a~spin-1 boson having a mass of a few GeV and interactions with both quarks and
leptons has been proposed to explain the measured value of the muon $g$$-$2 and the NuTeV
anomaly simultaneously~\cite{Gninenko:2001hx,Boehm:2004uq}, although the latter may now be
explicable by taking into account the appropriate nuclear
effects~\cite{Thomas:2011cm,erler-langacker}.
As another example, an ${\cal O}$(MeV) spin-1 boson which couples to dark matter as well
as leptons may be the cause of the observed 511-keV emission from the bulge of our
galaxy~\cite{Hooper:2007jr,Fayet:2007ua}.
If its mass is at the GeV level, such a particle may be associated with the unexpected
excess of positrons seen in cosmic rays, potentially attributable to dark-matter
annihilation~\cite{Foot:1994vd}.
In the context of hyperon decays, a~spin-1 boson with mass around $0.2$\,GeV,
flavor-changing couplings to quarks, and a primary decay channel into $\mu^+\mu^-$
can account for the three anomalous events of \,$\Sigma^+\to p\mu^+\mu^-$\, detected in
the HyperCP experiment several years ago~\cite{hypercp}.
Lastly, a~spin-1 particle lighter than the $b$~quark could be responsible~\cite{Oh:2010vc} for
the unexpectedly sizable like-sign dimuon charge asymmetry in semileptonic $b$-hadron
decays recently reported by the D{\O} Collaboration~\cite{Abazov:2010hv}.
Although in these few instances the spin-1 particles tend to have suppressed couplings to SM
particles, it is possible to test their existence or effects in future high-precision
experiments~\cite{Hooper:2007jr,Foot:1994vd,hypercp,Oh:2010vc,Pospelov:2008zw,Reece:2009un}.

In the present paper, we explore the possibility that a nonstandard spin-1 boson under 100\,GeV
is electrically neutral, carries no color, and has couplings to both the neutrinos and electron.
Consequently, it will affect processes that involve at least these leptons. In particular,
we will focus on such processes for which plenty of experimental data are available.
In our study, the new particle, to which we refer as the $X$ boson, is not necessarily a gauge
boson.  Therefore, its couplings to the leptons are kept sufficiently general for
a model-independent analysis.  The results of our analysis can be readily applied to
the specific case where $X$ is a gauge boson or any model with definite couplings of~$X$.
For simplicity, we also assume that the $X$ boson does not mix with the SM gauge bosons,
{\it i.e.}, $Z$ and $\gamma$.
As alluded to earlier, most previous NSI studies concentrate on the scenario of heavy new
particles.  As far as we know, the low-mass effects of $X$ on the determination of its
couplings have not been studied in detail before.  When the mass of $X$ is close to the momenta
exchanged in a scattering process, both the exchanged momenta and the $X$-boson mass
(and even its total decay width) have to be kept in the calculations.
This work is complementary to analyses on neutrino NSI due to new physics above the electroweak
scale~({\it e.g.}, Refs.\,\cite{Berezhiani:2001rs,Barranco:2007ej,Forero:2011zz}).

The structure of this paper is organized as follows.
In the next section, we write down an interaction Lagrangian for $X$ with the leptons
and subsequently describe the (anti)neutrino-electron and \,$e^+e^-\to\nu\bar\nu\gamma$\,
scattering processes that will be used to constrain possible values of
the $X$ couplings to the leptons.
From Section~\ref{nue} to Section~\ref{separate}, we concentrate on the flavor-conserving
interactions.  In Sections~\ref{nue} and~\ref{anue}, we extract constraints on the $X$
couplings to the electron and to the electron neutrino and antineutrino from the low-energy
\,$\nu_e^{}e\to\nu e$\, and \,$\bar\nu_e^{}e\to\bar\nu e$\, data, respectively.
A~combined result from the two sets of data is presented at the end of Section~\ref{anue}.
Section~\ref{numue} deals with the bounds on the $X$ couplings to the electron and
to the muon neutrinos based on the CHARM-II data.
In Section~\ref{ee2nng}, the data on \,$e^+e^-\to\nu\bar\nu\gamma$\, cross-section collected
by the ALEPH, DELPHI, L3, and OPAL Collaborations are employed to restrict the leptonic
couplings of~$X$, with several illustrative choices of its mass and total decay width.
We also discuss the complementarity of the (anti)neutrino-electron and
\,$e^+e^-\to\nu\bar\nu\gamma$\, measurements in probing these couplings.
Since the $X$ contributions to these observables always involve the products of its
respective couplings to the electron and neutrinos, the resulting constraints also apply only
to the products, instead of the individual couplings.
It is therefore of interest to determine constraints on the separate couplings making use of
other experimental information, which is also available.
We pursue this in Section~\ref{separate} for the flavor-conserving couplings employing
$e^+e^-$ collision data at the $Z$-pole, the measured anomalous magnetic moment of
the electron, and the results of searches for a nonstandard spin-1 particle at fixed-target
and beam-dump experiments.
Comparing the various results would allow us to see which observables are most sensitive to
the different $X$ couplings.
Finally, in Section~\ref{fcc} we address constraints on the couplings for the flavor-changing
$X$-neutrino interactions from the same sets of experimental data utilized in
Sections~\ref{nue}-\ref{ee2nng}.  Our findings are summarized in Section~\ref{sec:summary}.
Some longer formulas are collected in an~appendix.

%%%%%%%%%%%%%%%%%%%%%%%%%%%%%%%%%%%%%%%%%%%%%%%%%%
\section{Interactions and cross sections\label{interactions}}
%%%%%%%%%%%%%%%%%%%%%%%%%%%%%%%%%%%%%%%%%%%%%%%%%%

The Lagrangian describing the effective interactions of $X$ with the neutrinos,~$\nu_i^{}$,
and electron,~$e$, can take the form
\begin{eqnarray} \label{LX}
{\cal L}_X^{}  \,\,=\,\,
-g_{\nu_i^{}\nu_j^{}}^{}\,\bar\nu_i^{}\gamma^\beta P_L^{}\nu_j^{}\,X_\beta^{} \,-\,
\bar e\gamma^\beta\bigl(g_{Le}^{}P_L^{}+g_{Re}^{}P_R^{}\bigr)e\,X_\beta^{} ~,
\end{eqnarray}
where summation over \,$i,j=e,\mu,\tau$\, is implied, we have allowed for the possibility of
$X$-induced neutrino flavor-change, and \,$P_{L,R}^{}=\frac{1}{2}(1\mp\gamma_5^{})$.\,
Since presently there is still no compelling evidence for the existence of predominantly
right-handed neutrinos~\cite{nupdg}, we have neglected their potential couplings to~$X$.
We also have not included terms involving the muon or tau, as the electron is the only
charged lepton taking part in the reactions we will study.
The Hermiticity of ${\cal L}_X^{}$ implies that \,$g_{\nu_i\nu_j}^{}=g_{\nu_j\nu_i}^*$\,
and that $g_{Le,Re}^{}$ are real.
In our model-independent approach, we assume that these parameters are free and can be
family nonuniversal.
We further assume that additional coupling constants which $X$ may have parametrizing its
interactions, flavor-conserving and/or flavor-violating, with other fermions already satisfy
the experimental constraints to which the couplings are subject, but which we do not address
in this paper.

In the SM, neutrino-electron interactions proceed from diagrams with
the $W$ and $Z$ bosons exchanged between the fermions.
The relevant Lagrangian is given by
\begin{eqnarray} & \displaystyle
{\cal L}_{\rm SM}^{} \,\,=\,\,
-\frac{g}{\sqrt2}\bigl(\bar\nu_e^{}\gamma^\beta P_L^{}e\,W_\beta^+ +{\rm H.c.}\bigr) \,-\,
\frac{g}{2c_{\rm w}^{}}\,\bar\nu_i^{}\gamma^\beta P_L^{}\nu_i^{}\,Z_\beta^{} \,-\,
\frac{g}{c_{\rm w}^{}}\,\bar e\gamma^\beta
\bigl(\bar g_L^{}P_L^{}+\bar g_R^{}P_R^{}\bigr)e\,Z_\beta^{} ~, ~~~~
& \\ &
\bar g_L^{} \,\,=\,\, -\frac{1}{2} + s_{\rm w}^2 ~, \hspace{5ex}
\bar g_R^{} \,\,=\,\, s_{\rm w}^2 \,\,=\,\, \sin^2\theta_W^{} ~, \hspace{5ex}
c_{\rm w}^{} \,\,=\,\, \cos\theta_W^{} ~, & \label{Lsm}
\end{eqnarray}
where as usual $g$ is the weak coupling constant and $\theta_{W}^{}$ the Weinberg angle.

One can place bounds on the products of $X$ couplings to the neutrino and electron in
Eq.\,(\ref{LX}) from the cross sections of \,$\nu e\to\nu e$\, and
\,$\bar\nu e\to\bar\nu e$\, scattering which have been determined in a~number of low-energy
experiments~\cite{Allen:1992qe,Auerbach:2001wg,Reines:1976pv,krasno,rovno,
Daraktchieva:2003dr,texono,Beyer:1994mc,Vilain:1994qy,Boehm:1987fc}.
The accumulated data are generally consistent with SM expectations,
but there is room left for new physics.

In the SM, the amplitude for \,$\nu_e^{}e^-\to\nu_e^{}e^-$\, at tree level comes from
$u$-channel $W$-mediated and $t$-channel $Z$-mediated diagrams, while for
\,$\nu_\mu^{}e^-\to\nu_\mu^{}e^-$\, the $W$ contribution is absent~\cite{'tHooft:1971ht}.
For these processes, the $X$ interactions in Eq.\,(\ref{LX}) can induce $t$-channel diagrams.
The latter type of $X$-mediated diagram is the only contribution at leading order to
\,$\nu_i^{}e^-\to\nu_j^{}e^-$\, for \,$j\neq i$\, in the absence of other nonstandard mechanisms.
Since the final neutrino in the $\nu e$ scattering experiments is not detected,
any one of the three light-neutrino flavors can occur in the final state.
It follows that for \,$\nu_i^{}e^-\to\nu e^-$\, and \,$i=e$ or $\mu$\, we have
the differential cross-section
\begin{eqnarray}
\frac{d\sigma_{\nu_i^{}e}^{}}{dT} \,\,=\,\,
\frac{1}{32\pi E_{\nu\,}^2m_e^{}}\,\sum_{j=e,\mu,\tau}
\overline{\bigl|{\cal M}_{\nu_i^{}e\to\nu_j^{}e}\bigr|^2} ~,
\end{eqnarray}
where $E_\nu$ and $T$ denote, respectively, the energy of the incident neutrino and the kinetic
energy of the recoiling electron both in the laboratory frame, $m_e^{}$ is the electron mass,
and the general expressions for the squared amplitudes can be found
in Eqs.~(\ref{M2nuee})-(\ref{M2nule}) in the Appendix.

In the case that the momentum transfers in the scattering are small compared to
the $W$ and $X$ masses, we can write approximately
\begin{eqnarray} \label{dcs/dT}
\frac{d\sigma_{\nu_i^{}e}^{}}{dT} \,\,=\,\, \frac{d\sigma_{\nu_i^{}e}^{\rm FD}}{dT} \,+\,
\frac{d\sigma_{\nu_i^{}e}^{\rm FC}}{dT} ~,
\end{eqnarray}
\begin{eqnarray} \label{fd}
\frac{d\sigma_{\nu_i^{}e}^{\rm FD}}{dT} &=&
\frac{2 G_{\rm F}^2 m_e^{}}{\pi} \left[
\Biggl(\omega+\bar g_L^{}+\frac{L_{ii}^{}}{2\sqrt2\,G_{\rm F\,}^{}m_X^2}\Biggr)^{\!\!2} +
\Biggl(\bar g_R^{}+\frac{R_{ii}^{}}{2\sqrt2\,G_{\rm F\,}^{}m_X^2}\Biggr)^{\!\!2}
\biggl(1-\frac{T}{E_\nu}\biggr)^{\!\!2} \right.
\nonumber \\ && \left. \hspace{9ex} -\;
\Biggl(\omega+\bar g_L^{}+\frac{L_{ii}^{}}{2\sqrt2\,G_{\rm F\,}^{}m_X^2}\Biggr)
\Biggl(\bar g_R^{}+\frac{R_{ii}^{}}{2\sqrt2\,G_{\rm F\,}^{}m_X^2}\Biggr)
\frac{m_e^{}T}{E_\nu^2} \right] ,
\end{eqnarray}
\begin{eqnarray} \label{fc}
\frac{d\sigma_{\nu_i^{}e}^{\rm FC}}{dT} \,\,=\,\,
\frac{m_e^{}}{4\pi m_X^4} \sum_{j\neq i}
\Biggl[ \bigl|L_{ji}\bigr|^2+\bigl|R_{ji}\bigr|^2\biggl(1-\frac{T}{E_\nu}\biggr)^{\!\!2}
- L_{ij}R_{ji}\,\frac{m_e^{}T}{E_\nu^2} \Biggr] ~,
\end{eqnarray}
where the two parts in Eq.\,(\ref{dcs/dT}) arise from flavor-diagonal (FD) and
flavor-changing (FC) interactions, respectively,
\,$G_{\rm F}^{}=g^2/\bigl(32m_W^4\bigr){}^{1/2}$\, as usual,
\,$\omega=1\,(0)$\, if \,$i=e\,(\mu)$,\,
and \,${\sf C}_{ij}^{}=g_{\nu_i\nu_j}^{}g_{{\sf C}e}^{}$\, for \,${\sf C}=L,R$,\, implying
that \,${\sf C}_{ij}^*={\sf C}_{ji}^{}$\,
and~\,${\rm Re}\bigl(L_{ji}^*R_{ji}^{}\bigr)=L_{ij}R_{ji}$.\,
In the \,$L_{ii}=R_{ii}=0$\, limit, Eq.\,(\ref{fd}) reproduces the well-known SM
contribution~\cite{'tHooft:1971ht}.
For \,$i=e$\, and \,$E_\nu^{}\gg m_e^{}$,\, we then arrive at
\begin{eqnarray} \label{csnue}
\sigma_{\nu_e^{}e}^{} &=&
\frac{2 G_{\rm F}^2 E_\nu^{} m_e^{}}{\pi} \left[
\Biggl(1+\bar g_L^{}+\frac{L_{ee}^{}}{2\sqrt2\,G_{\rm F\,}^{}m_X^2}\Biggr)^{\!\!2} + \frac{1}{3}
\Biggl(\bar g_R^{}+\frac{R_{ee}^{}}{2\sqrt2\,G_{\rm F\,}^{}m_X^2}\Biggr)^{\!\!2} \right]
\nonumber \\ && +\;
\frac{E_\nu^{}m_e^{}}{4\pi\,m_X^4}
\Biggl( \bigl|L_{\mu e}\bigr|^2+\frac{\bigl|R_{\mu e}\bigr|^2}{3} \,+\, (\mu\to\tau) \Biggr)
\end{eqnarray}
after integration over the $T$ range in Eq.\,(\ref{Trange}) and keeping terms to first order
in~$m_e^{}$.

If $m_X^{}$ is not large compared to the momentum transfer, one needs to employ the general
expressions in Eqs.~(\ref{M2nuee})-(\ref{M2nule}) to calculate
the cross sections~$\sigma_{\nu_i^{}e}$, but the approximations such as made in the previous
paragraph are still applicable to the SM part for momenta much smaller than~$m_W^{}$.
Moreover, for $\nu_e^{}e$ scattering with incident neutrinos having been produced in $\mu^+$
decays at rest and therefore not being monoenergetic, one has to integrate
$\sigma_{\nu_e^{}e}$ over the appropriate $\nu_e^{}$ spectrum~\cite{Kayser:1979mj}.
This results in the flux-averaged cross-section~\cite{Allen:1992qe}
\begin{eqnarray} \label{facs}
\bar\sigma_{\nu_e^{}e}^{} \,\,=\,\,
\int_0^{E_\nu^{\rm max}} dE_\nu^{}\;\phi_{\nu_e\!}^{}(E_\nu)\;\sigma_{\nu_e^{}e}^{} ~,
\end{eqnarray}
where the limits span the $\nu_e^{}$ energy range in $\mu^+$ decay,
\,$E_\nu^{\rm max}=\bigl(m_\mu^2-m_e^2\bigr)/(2m_\mu)\simeq52.8$\,MeV\, with
the $\nu$ masses neglected, and the spectrum is given by~\cite{Kayser:1979mj}
\,$\phi_{\nu_e}(E_\nu)=12\bigl(E_\nu^{\rm max}-E_\nu^{}\bigr)E_\nu^2/(E_\nu^{\rm max})^4$,\,
which is normalized to unity.

In the \,$\bar\nu_e^{}e^-\to\bar\nu e^-$\, processes of interest, the source of
the incident antineutrinos is a~nuclear reactor and hence they do not share the same energy.
The cross section then again needs to be integrated over the reactor antineutrino
spectrum~\cite{Forero:2011zz,Kayser:1979mj},
\begin{eqnarray} \label{nubar-e}
\bar\sigma_{\bar\nu_e^{}e}^{} \,\,=\,\,
\int_{T_{\rm min}}^{T_{\rm max}}dT\int_{E_{\bar\nu}^{\rm min}}^{E_{\bar\nu}^{\rm max}}
dE_{\bar\nu}^{}\;\phi_{\bar\nu_e\!}^{}(E_{\bar\nu})\;\frac{d\sigma_{\bar\nu_e^{}e}^{}}{dT} ~,
\end{eqnarray}
where $T_{\rm min,max}$ denote the experimental cuts on the kinetic energy $T$ of the recoiling
electron in the lab frame, $E_{\bar\nu}^{\rm min}$ is a function of $T$ according
to~Eq.\,(\ref{Enumin}), and the spectrum, which extends essentially
to \,$E_{\bar\nu}^{\rm max}\sim10$\,MeV,\, is given by~\cite{Barranco:2007ej,Forero:2011zz}
\begin{eqnarray} \label{flux}
\phi_{\bar\nu_e\!}^{}(E_{\bar\nu}) \,\,=\,\, \sum_k a_k^{}\, S_k^{}(E_{\bar\nu}) ~,
\end{eqnarray}
the sum of the spectra $S_k(E_{\bar\nu})$ from isotopes $k$ with fractional
contributions~$a_k^{}$.
The differential cross-section \,$d\sigma_{\bar\nu_e^{}e}/dT$\, for $m_{W,X}^{}$ large compared to
the total energy in this scattering can be derived from Eqs.~(\ref{dcs/dT})-(\ref{fc}) by making
the interchanges \,$1+\bar g_L^{}\leftrightarrow\bar g_R^{}$\,
and~\,$L_{ij}\leftrightarrow R_{ij}$.\,
If $m_X^{}$ is not much greater than the momentum transfer in this reaction, one needs to use
the \,$\bar\nu_e^{}e^-\to\bar\nu e^-$\, counterparts of Eqs.~(\ref{M2nuee})-(\ref{M2nule})
in evaluating the cross sections.

Additional bounds on the $X$ couplings to the leptons are available from
\,$e^+e^-\to\nu\bar\nu\gamma$\, scattering, which has been observed
at~LEP~\cite{Buskulic:1996hw,Barate:1997ue,Barate:1998ci,Heister:2002ut,Abreu:2000vk,
Abdallah:2003np,Acciarri:1997dq,Acciarri:1998hb,Acciarri:1999kp,Ackerstaff:1997ze,
Abbiendi:1998yu,Abbiendi:2000hh}.
The cross section of this process has been computed in the literature for
the SM~\cite{Ma:1978zm,Berends:1987zz} as well as its extensions containing extra
charged and neutral gauge bosons~\cite{Godfrey:2000hc}.
In the SM the amplitude at tree level is generated by five diagrams, three of which are
mediated by the $W$ and two by the~$Z$.
The $X$ contributions are similar in form to the $Z$ diagrams.
Our calculations including the $X$ contributions agree with the earlier
results~\cite{Berends:1987zz,Godfrey:2000hc}.
The cross section can be written as
\begin{eqnarray} \label{cs_ee2nng}
\sigma_{e\bar e\to\nu\bar\nu\gamma}^{} \,\,=\,\,
\frac{1}{2(4\pi)^4\,(p_{e^+}+p_{e^-})^2}
\int dE_\gamma^{}\,E_\gamma^{}\;d\bigl(\cos\theta_\gamma^{}\bigr)\;d\bar\Omega_\nu^{}
\sum_{i,j=e,\mu,\tau}
\overline{\bigl|{\cal M}_{e\bar e\to\nu_i^{}\bar\nu_j^{}\gamma}^{}\bigr|^2} ~,
\end{eqnarray}
where $E_\gamma^{}$ and $\theta_\gamma^{}$ are the photon energy and angle with respect to
the $e^+$ or $e^-$ beam direction in the $e^+e^-$ center-of-mass frame,
$\bar\Omega_\nu^{}$ denotes the solid angle of either $\nu$ or $\bar\nu$ in the $\nu\bar\nu$
center-of-mass frame, and the formulas for the squared amplitudes are given
in~Eqs.\,(\ref{M2ee2nng})-(\ref{M2ee2nng''}).
Our numerical analysis starts in the next section.

%%%%%%%%%%%%%%%%%%%%%%%%%%%%%%%%%%%%%%%%%%%%%%%%%%
\section{Constraints from \,$\bm{\nu_e^{}e\to\nu e}$\label{nue}}
%%%%%%%%%%%%%%%%%%%%%%%%%%%%%%%%%%%%%%%%%%%%%%%%%%

The latest data on the cross section of \,$\nu_e^{}e^-\to\nu e^-$\, have been acquired in
the E225 experiment at LAMPF~\cite{Allen:1992qe} and the LSND experiment~\cite{Auerbach:2001wg}.
They measured the flux-averaged cross-sections
\,$\bar\sigma_{\nu_e e}^{\rm exp}=(3.18\pm0.56)\times10^{-43}{\rm\,cm}^2$\, and
\,$\bar\sigma_{\nu_e e}^{\rm exp}=(3.19\pm0.48)\times10^{-43}{\rm\,cm}^2$,\,
respectively, corresponding to
\,$\sigma_{\nu_e e}^{\rm exp}=(10.0\pm1.8)\times10^{-45}{\rm\,cm}^2\bar E_\nu^{}/\rm MeV$\, and
\,$\sigma_{\nu_e e}^{\rm exp}=
(10.1\pm1.5)\times10^{-45}{\rm\,cm}^2\bar E_\nu^{}/\rm MeV$\,~\cite{Allen:1992qe,Auerbach:2001wg}
with flux-averaged energy \,$\bar E_\nu\simeq31.7$\,MeV,\,
the statistical and systematic errors of each having been combined in quadrature.
The SM prediction is
\,$\sigma_{\nu_e e}^{\rm SM}=9.3\times10^{-45}{\rm\,cm}^2E_\nu^{}/\rm MeV$~\cite{Auerbach:2001wg},
which translates into~\,$\bar\sigma_{\nu_e e}^{\rm SM}=2.95\times10^{-43}{\rm\,cm}^2$.\,
Performing unconstrained averaging of the two measurements following the Particle Data Group
prescription~\cite{pdg} yields
\,$\bar\sigma_{\nu_e e}^{\rm exp}=(3.19\pm0.37)\times10^{-43}{\rm\,cm}^2$,\, leading
to~\,$\sigma_{\nu_e e}^{\rm exp}=(10.1\pm1.2)\times10^{-45}{\rm\,cm}^2\bar E_\nu^{}/\rm MeV$.\,

To explore the constraints on the $X$ couplings to the leptons from this average value
of~$\bar\sigma_{\nu_e e}^{\rm exp}$, we adopt its 1.64-sigma [90\% confidence level~(CL)]
limits and isolate the $X$ contribution, including its interference with the SM amplitude, by
subtracting out the SM cross-section, $\bar\sigma_{\nu_e e}^{\rm SM}$, quoted above.
In numerical calculations, we will take \,$G_{\rm F}^{}=1.166\times10^{-5}{\rm~GeV}^{-2}$\,
and~\,$\sin^2\theta_W^{}=0.23$,\, unless otherwise stated.

\begin{figure}[t]
\includegraphics[width=73mm]{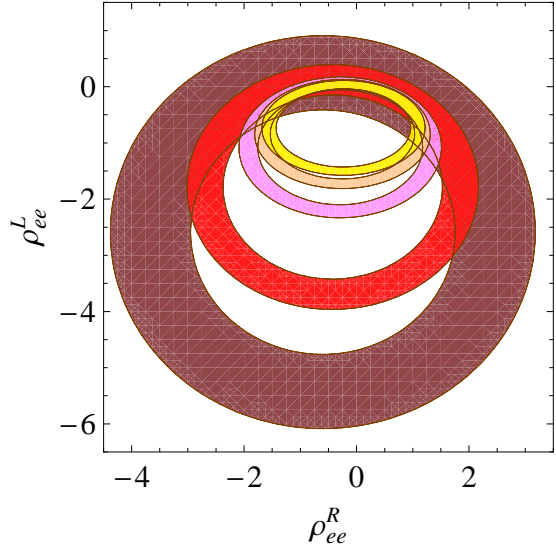}\vspace*{-1ex}
\caption{Values of $\rho_{ee}^{L}$ and $\rho_{ee}^{R}$ subject to constraints from LAMPF and
LSND data on~\,$\nu_e^{}e\to\nu e$\, scattering for, from largest to smallest
rings,~\,$m_X^{}=1,2,5,10,50$~MeV.\,  The yellow (most lightly-shaded) ring,
for~\,$m_X^{}=50$\,MeV,\, is virtually identical to that for any
other~\,$m_X^{}>40$\,MeV.\label{lampf+lsnd}}
\end{figure}

We address first the flavor-conserving couplings, turning off the flavor-changing ones
in this and the next four sections.
Assuming that the coupling products $L_{ee}$ and $R_{ee}$ occur at the same time in the $X$
contribution to the flux-averaged cross-section in Eq.\,(\ref{facs}), with the squared amplitude
given by that in~Eq.\,(\ref{MX2nuee}), we try various values of the $X$ mass.
It turns out that the allowed ranges of the ratios of these parameters to the squared $X$-mass
become less and less dependent on $m_X^{}$ fairly quickly if it exceeds
\,{\small$\sim$\,}40\,MeV\, and that below this value the effect of the low $m_X^{}$ on
the allowed regions increasingly manifests itself as $m_X^{}$ decreases.
In particular, the restrictions on the ratios grow weaker as the lighter masses get lower.
We illustrate all this in~Fig.\,\ref{lampf+lsnd} for several examples of $m_X^{}$ values,
where $\rho_{ee}^{L,R}$ are the ratios normalized by $2\sqrt2\,G_{\rm F}^{}$ according to
the general definition
\begin{eqnarray} \label{rho}
\rho_{ij}^{\sf C} \,\,=\,\, \frac{{\sf C}_{ij}^{}}{2\sqrt2\,G_{\rm F}^{}m_X^2} ~, \hspace{5ex}
{\sf C}_{ij}^{} \,\,=\,\, g_{\nu_i^{}\nu_j^{}\,}^{}g_{{\sf C}e}^{} ~,
\hspace{5ex} {\sf C}^{} \,\,=\,\, L,R ~.
\end{eqnarray}
It is worth noting that, since in this reaction the magnitude of the momentum exchange in
the $X$ diagram is less than
\,$|t|_{\rm max}^{1/2}\simeq(2E_\nu^{\rm max}m_e^{})^{1/2}\simeq7.4$\,MeV,\, the application
of the approximate formula in Eq.\,(\ref{csnue}) for \,$m_X^{}<40$\,MeV\, would entail errors
of more than \,$|t|_{\rm max}^{}/m_X^2\sim3\%$.\,

%%%%%%%%%%%%%%%%%%%%%%%%%%%%%%%%%%%%%%%%%%%%%%%%%%
\section{Constraints from \,$\bm{\bar\nu_e^{}e\to\bar\nu e}$\label{anue}}
%%%%%%%%%%%%%%%%%%%%%%%%%%%%%%%%%%%%%%%%%%%%%%%%%%

\begin{table}[b]
\caption{Experimental results on \,$\bar\nu_e^{}e^-\to\bar\nu e^-$\, scattering
cross section $\bar\sigma$ or event rate $R$.} \footnotesize
\begin{tabular}{|c|cc|}
\hline
Experiment & $T$ (MeV) & Measurement \\
\hline\hline
\, Savannah River~\cite{Reines:1976pv} \, & 1.5\,-\,3.0 & ~
$\bar\sigma=(0.87\pm0.25)\bar\sigma_{V\mbox{-}\!A}^{}$ \\
& 3.0\,-\,4.5 & $\bar\sigma=(1.70\pm0.44)\bar\sigma_{V\mbox{-}\!A}^{}$ \\
Krasnoyarsk~\cite{krasno} & \, 3.150\,-\,5.175 \, &
$\bar\sigma=(4.5\pm2.4)\times10^{-46}\rm\,cm^2$/fission \\
Rovno~\cite{rovno} & 0.6\,-\,2.0 & \, $\bar\sigma=(1.26\pm0.62)\times10^{-44}\rm\,cm^2$/fission \, \\
MUNU~\cite{Daraktchieva:2003dr} & 0.7\,-\,2.0 & $R=(1.05\pm0.35)R_{\rm SM}$ \\
Texono~\cite{texono} & 3.0\,-\,8.0 & $R=(1.08\pm0.26)R_{\rm SM_{\vphantom{|}}}$ \\
\hline
\end{tabular} \label{antinuexp}
\end{table}

The cross section of \,$\bar\nu_e^{}e^-\to\bar\nu e^-$\, has been evaluated in several experiments
at nuclear power plants.   The data on its flux-averaged value
\,$\bar\sigma=\bar\sigma_{\bar\nu_e e}^{\rm exp}$\, or the corresponding event rate $R$,
along with their ranges of the final electron's kinetic energy $T$, are
listed in~Table~\ref{antinuexp}.\footnote{\baselineskip=14pt%
In the Savannah River entries, $\bar\sigma_{V\mbox{-}\!A}^{}$ is the corresponding cross section
in the SM from the $W$-mediated diagram alone.  The MUNU number for $R/R_{\rm SM}$ has been
obtained from the observed \,$R=1.07\pm0.34$~counts/day\, and expected
\,$R_{\rm SM}=1.02\pm0.10$~counts/day~\cite{Daraktchieva:2003dr}.}
To extract the $X$ couplings permitted by these data, we adopt again the 90\%-CL ranges of
the experimental numbers and subtract out from Eq.\,(\ref{nubar-e}) the pure SM part given by
the usual approximation~\cite{'tHooft:1971ht}
\begin{eqnarray} \label{dcssm'}
\frac{d\sigma_{\bar\nu_e^{}e}^{\rm SM}}{dT} \,\,=\,\, \frac{2 G_{\rm F}^2 m_e^{}}{\pi} \Biggl[
\bigl(1+\bar g_L^{}\bigr)^{\!2}\biggl(1-\frac{T}{E_{\bar\nu}}\biggr)^{\!\!2} \,+\, \bar g_R^2
\,-\, \bigl(1+\bar g_L^{}\bigr)\bar g_R^{}\,\frac{m_e^{}T}{E_{\bar\nu}^2} \Biggr]
\end{eqnarray}
appropriate in the \,$s\ll m_W^2$\, case.
For the antineutrino spectrum in~Eq.\,(\ref{flux}), the relevant isotopes are
\,$k={}^{235}{\rm U},{}^{238}{\rm U},{}^{239}{\rm Pu},{}^{241}{\rm Pu}$,\, their relative
contributions are taken to be the typical average (over an annual reactor cycle)
values \,$a_k^{}=0.54,0.07,0.33,0.06$\,~\cite{Daraktchieva:2003dr,Boehm:1987fc}, respectively,
and we employ the $S_k(E_{\bar\nu})$ parametrization provided in~Ref.\,\cite{Mueller:2011nm}.

\begin{figure}[b] \vspace{1ex}
\includegraphics[width=77mm]{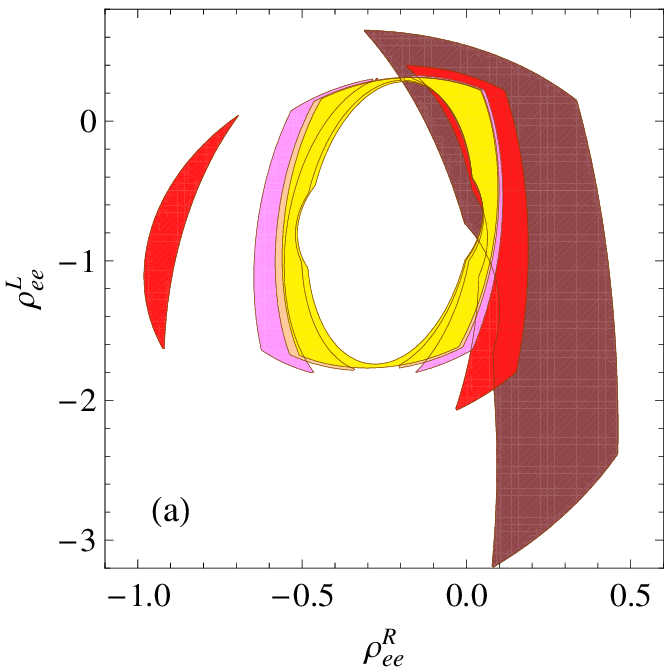} \, \, \,
\includegraphics[width=77mm]{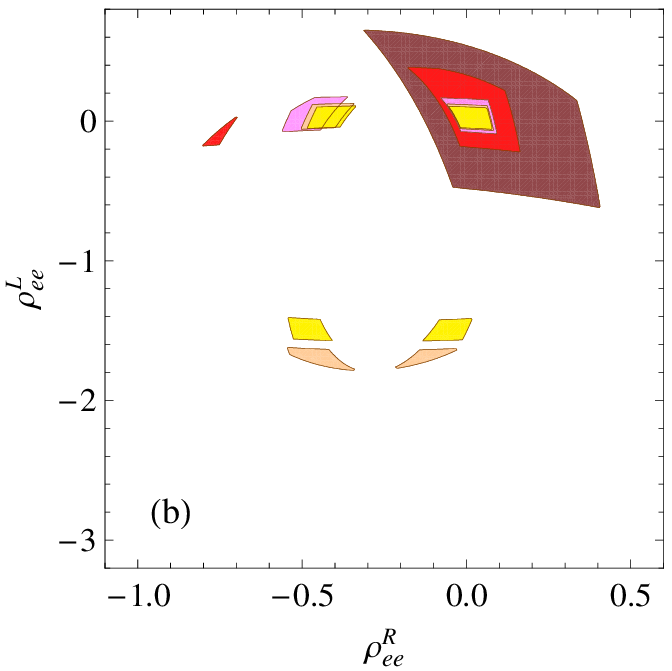}\vspace*{-1ex}
\caption{(a) Values of $\rho_{ee}^{L,R}$ subject to constraints from
\,$\bar\nu_e^{}e\to\bar\nu e$\, data for \,$m_X^{}=1$\,MeV\,(brown, darkest colored),
2\,MeV\,(red), 5\,MeV\,(magenta), 10\,MeV\,(orange), 50\,MeV\,(yellow, most lightly-shaded).\,
The (yellow) area for \,$m_X^{}=50$\,MeV\, is virtually identical to that for any
other~\,$m_X^{}>25$\,MeV.\,
(b)~Overlaps between the allowed regions in (a) and Fig.\,\ref{lampf+lsnd}.\label{ll+reactors}}
\end{figure}

With the two flavor-conserving parameters, $L_{ee}$ and $R_{ee}$, being present simultaneously
as before, we scan the parameter space to observe that the low-mass effect of $X$ on
the allowed regions of its (squared) coupling-to-mass ratios begins to appear strikingly as
$m_X^{}$ goes below {\small\,$\sim$\,}25\,MeV\, and that the bounds tend to become weaker as
the mass gets lower, similar to the $\nu_e^{}e$ case.
This pattern is depicted in Fig.\,\ref{ll+reactors}(a) with some illustrative values of~$m_X^{}$.
Since in this reaction the momentum exchange basically has a size of less than
\,$s_{\rm max}^{1/2}=\bigl(2E_{\bar\nu}^{\rm max}m_e^{}+m_e^2\bigr){}^{1/2}\simeq3.2$\,MeV,\,
the use of the approximate formula of the $\bar\nu_e e$ counterpart of Eq.\,(\ref{fd}) for
\,$m_X^{}<25$\,MeV\, would expectedly generate errors of more
than~\,$s_{\rm max}^{}/m_X^2\sim2\%$.\,

We can now combine the constraints from the \,$\nu_e^{}e\to\nu e$\, and
\,$\bar\nu_e^{}e\to\bar\nu e$\, measurements above.
We show the overlap areas satisfying the two sets of data in~Fig.\,\ref{ll+reactors}(b).
It is clear that the joint constraints reduce the $\rho_{ee}^{L,R}$ ranges significantly.
In this graph, their extreme values specifically are
\,$\bigl(\rho_{ee,\rm min}^L,\rho_{ee,\rm max}^L\bigr)=
(-0.62,0.65),(-0.22,0.39),(-0.09,0.17),(-1.79,0.12),(-1.57,0.11)$\,
and
\,$\bigl(\rho_{ee,\rm min}^R,\rho_{ee,\rm max}^R\bigr)=
(-0.31,0.41),(-0.81,0.16),(-0.56,0.09),(-0.55,0.08),(-0.54,0.08)$\,
for \,$m_X^{}=1,2,5,10,50$~MeV,\, respectively, corresponding to the upper limits of
$|L_{ee}|^{1/2}$ and $|R_{ee}|^{1/2}$ varying roughly from \,$4\times10^{-6}$\,
to~\,$4\times10^{-4}$.\,
If instead \,$m_X^{}=1,5,100$~GeV,\, the largest limits would be {\small\,$\sim$\,}0.007,\,0.04,\,0.7,\,
respectively, from the yellow (most lightly-shaded) areas.
Thus, although the low-$m_X^{}$ effect on the allowed $\rho_{ee}^{L,R}$ areas seen in the earlier
figures more or less persists here, the products of $X$ couplings to the electron neutrino
and electron still undergo increasing restraints from these data as $m_X^{}$ decreases.
We should mention that the yellow (most lightly-shaded) regions in Fig.\,\ref{ll+reactors}(b) for
\,$m_X^{}\ge50$\,MeV\, are comparable to their counterparts resulting from the model-independent
analyses in Refs.\,\cite{Barranco:2007ej,Forero:2011zz} on nonstandard neutrino-electron
interactions due to new physics above the electroweak scale.

\begin{figure}[b]
\includegraphics[width=85mm]{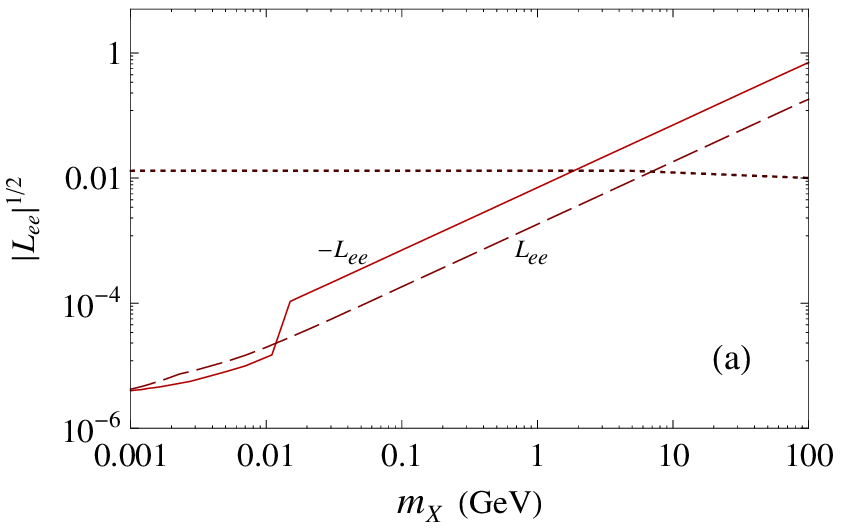} ~
\includegraphics[width=85mm]{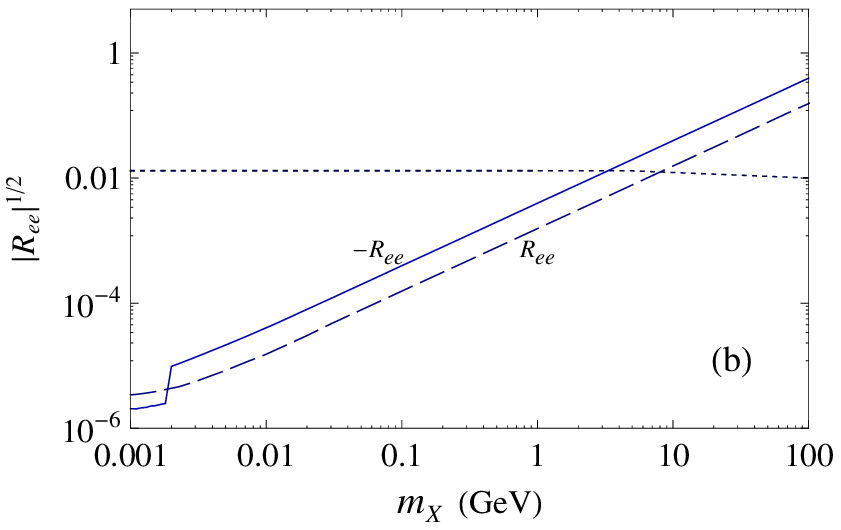} \vspace{-1ex} \\
\includegraphics[width=85mm]{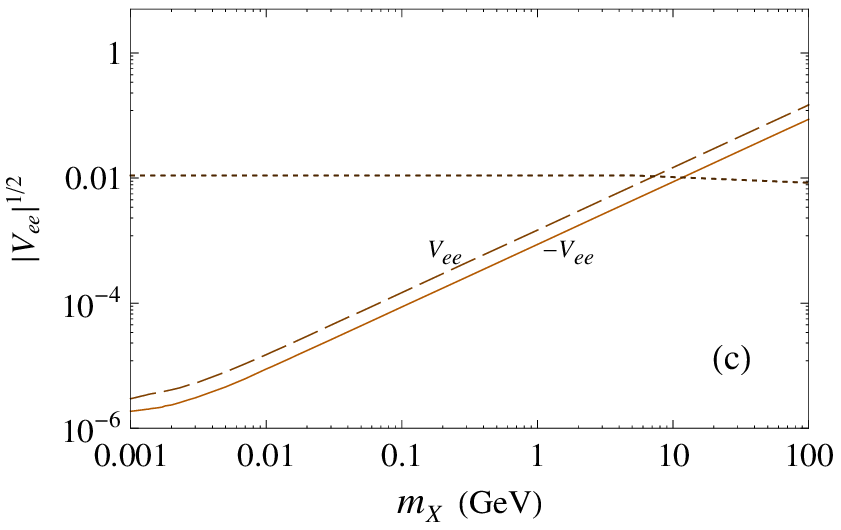} ~
\includegraphics[width=85mm]{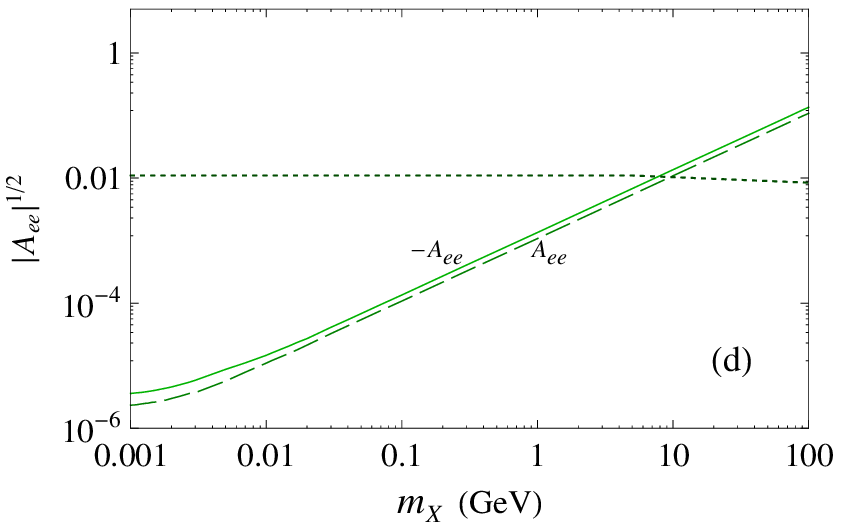} \vspace{-1ex}
\caption{Upper limits on (a)~$|L_{ee}|^{1/2}$ and (b)~$|R_{ee}|^{1/2}$ versus $X$
mass from the extreme values of $\pm L_{ee}$ and $\pm R_{ee}$, respectively, allowed by
\,$\nu_e^{}e\to\nu e$\, and \,$\bar\nu_e^{}e\to\bar\nu e$\, data only (solid and dashed curves)
or \,$e^+e^-\to\bar\nu\nu\gamma$\, data only (dotted curves), under the assumption that
the other coupling product is zero.
Also plotted are the corresponding limits for (c)~$V_{ee}=\frac{1}{2}(L_{ee}+R_{ee})$\,
and (d)~$A_{ee}=\frac{1}{2}(L_{ee}-R_{ee})$\, if only one of these combinations is
nonvanishing.\label{coupling-mass}}
\end{figure}

To paint a more complete picture about how the couplings are constrained by these
(anti)neutrino-electron scattering data,
we present in Fig. \ref{coupling-mass}(a,b) the upper limits on $|L_{ee}|^{1/2}$ and
$|R_{ee}|^{1/2}$ following from the allowed extreme values of $\pm L_{ee}$ and $\pm R_{ee}$,
respectively, over the $m_X^{}$ range of interest, assuming in each case that the other coupling
product is zero.
It is straightforward to see that the solid and dashed curves encompass the $|L_{ee}|^{1/2}$ or
$|R_{ee}|^{1/2}$ limits implied by the examples in~Fig.\,\ref{ll+reactors}(b).
The curves are also roughly compatible with the quoted numbers above for these quantities.
In Fig.~\ref{coupling-mass}(c,d), we display alternatively the bounds on the vector
and axial-vector combinations \,$V_{ee}=\frac{1}{2}(L_{ee}+R_{ee})$\, and
\,$A_{ee}=\frac{1}{2}(L_{ee}-R_{ee})$,\, respectively,\, extracted under the assumption again
that only one of them is nonzero.
All these special cases demonstrate that, as remarked in the last paragraph, the restrictions
become weaker as $m_X^{}$ gets larger.
This is unlike the behavior of the limits from \,$e^+e^-\to\nu\bar\nu\gamma$\, measurements,
which are represented by the dotted curves and will be discussed in Section\,\ref{ee2nng}.

%%%%%%%%%%%%%%%%%%%%%%%%%%%%%%%%%%%%%%%%%%%%%%%%%%
\section{Constraints from \,$\bm{\nu_\mu^{}e\to\nu e}$\label{numue}}
%%%%%%%%%%%%%%%%%%%%%%%%%%%%%%%%%%%%%%%%%%%%%%%%%%

\begin{figure}[b]
\includegraphics[width=70mm]{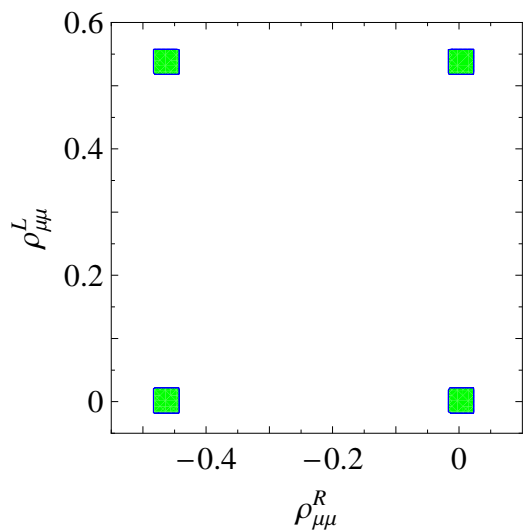}\vspace{-1ex}
\caption{Values of $\rho_{\mu\mu}^{L,R}$ allowed by CHARM-II data
on~\,$\nu_\mu^{}e\to\nu e$\, and \,$\bar\nu_\mu^{}e\to\bar\nu e$\, scattering
for~\,$m_X^{}\mbox{\small\,$\gtrsim$\,}1$\,GeV.\,\label{charmplot}}
\end{figure}

The scattering of a muon (anti)neutrino off an electron can probe the $X$ interactions
with them.
The most precise experiment on \,$\nu_\mu^{}e^-\to\nu e^-$\, and
\,$\bar\nu_\mu^{}e^-\to\bar\nu e^-$\, was carried out by the CHARM-II
Collaboration.
Their low-energy measurement~\cite{Vilain:1994qy}, based on the differential cross-section
\begin{eqnarray} \label{dcscharm}
\frac{d\sigma_{\nu_\mu^{}e}^{}}{dy} \,\,=\,\, \frac{G_{\rm F}^{2\,}E_\nu^{}m_e^{}}{2\pi}
\bigl[(g_V^{}+g_A^{})^2+(g_V^{}-g_A^{})^2(1-y)^2\bigr] ~,
\end{eqnarray}
where \,$y=T/E_\nu^{}$,\, and a similar expression for \,$\bar\nu_\mu^{}e^-\to\bar\nu e^-$\,
with $g_A^{}$ replaced by \,$-g_A^{}$,\, can be translated into
\begin{eqnarray} \label{charmdata}
\bigl(g_V^{\rm exp}+g_A^{\rm exp}\bigr)^{\!2} \,\,=\,\, 0.289\pm 0.026 ~, \hspace{5ex}
\bigl(g_V^{\rm exp}-g_A^{\rm exp}\bigr)^{\!2} \,\,=\,\, 0.219\pm 0.023 ~.
\end{eqnarray}
Comparing these cross-sections with the corresponding ones in Sec.\,\ref{interactions} and
assuming that $g_{V,A}^{\rm exp}$ consist of SM and $X$ terms, one can place bounds on
(the products of) the $X$ couplings to the muon (anti)neutrino and electron, depending on
the $X$ mass.  Adopting the 90\%-CL ranges of the numbers in Eq.\,(\ref{charmdata}) and
setting the flavor-changing couplings to zero, we obtain the allowed (green) regions of
$\rho_{\mu\mu}^{L}$ and $\rho_{\mu\mu}^{R}$ in Fig.\,\ref{charmplot} for transfer momenta
small compared to $m_X^{}$.
These results are comparable to their counterparts in the model-independent study of
Ref.\,\cite{Barranco:2007ej} on neutrino NSI due to new physics above
the electroweak scale.

For lower values of~$m_X^{}$, we are unable to derive bounds on these $\rho$ parameters
due to lack of the relevant information on the (anti)neutrino spectrum and
the flux-averaged cross-sections, unlike the $\nu_e^{}e$ and $\bar\nu_e^{}e$ cases.
Nevertheless, based on the findings of the preceding two sections, we can still draw the following
conclusion. Since in the CHARM-II experiment \,$T=3$-24\,GeV~\cite{Vilain:1994qy} leading to
the momentum exchange \,$|t|^{1/2}\le(2m_e^{}T_{\rm max})^{1/2}\simeq0.16$\,GeV,\, the green
(shaded) areas in Fig.\,\ref{charmplot} can be expected to be valid
for~\,$m_X^{}\mbox{\small\,$\gtrsim$\,}1$\,GeV\, with errors below~{\small\,$\sim$\,}2\%.
We can then infer that
\,$|L_{\mu\mu}|_{\rm max}^{1/2}\sim|R_{\mu\mu}|_{\rm max}^{1/2}\mbox{\small\;$\gtrsim$\;}0.004$\,
as $m_X^{}$ goes above~1\,GeV.

%%%%%%%%%%%%%%%%%%%%%%%%%%%%%%%%%%%%%%%%%%%%%%%%%%
\section{Constraints from \,$\bm{e^+e^-\to\nu\bar\nu\gamma}$\label{ee2nng}}
%%%%%%%%%%%%%%%%%%%%%%%%%%%%%%%%%%%%%%%%%%%%%%%%%%

The latest measurements of the \,$e^+e^-\to\nu\bar\nu\gamma$\, cross-section were
performed by the ALEPH, DELPHI, L3, and OPAL Collaborations at
LEP\,\cite{Buskulic:1996hw,Barate:1997ue,Barate:1998ci,Heister:2002ut,Abreu:2000vk,Abdallah:2003np,
Acciarri:1997dq,Acciarri:1998hb,Acciarri:1999kp,Ackerstaff:1997ze,Abbiendi:1998yu,
Abbiendi:2000hh} for various center-of-mass energies, $\hat s^{1/2}$, from about 130 to~207~GeV.
The acquired data along with the corresponding SM expectations are listed
in~Table~\ref{ee2nngdata}.
They allow us to impose the constraint
\,$\bigl|\hat\sigma_{\rm exp}^{}-\hat\sigma_{\rm SM}^{} -
\sigma_{e\bar e\to\nu\bar\nu\gamma}^X\bigr|\le
\bigl(\delta\sigma_{\rm exp}^2+\delta\sigma_{\rm SM}^2\bigr){}^{1/2}$,\,
where $\hat\sigma_{\rm exp,SM}^{}$ and $\delta\sigma_{\rm exp,SM}^{}$ are, respectively,
the central values and 90\%-CL uncertainties of the $\sigma_{\rm exp,SM}^{}$ numbers in
Table~\ref{ee2nngdata},\footnote{\baselineskip=14pt%
The 90\%-CL ranges of $\sigma_{\rm exp,SM^{\vphantom{0}}}$ in each one of the entries in
this table overlap, except for two in which the overlaps can occur at~2$\sigma$.
Assuming this to be due to statistical flukes, we use 2$\sigma$ uncertainties for these two
entries.\smallskip}
and $\sigma_{e\bar e\to\nu\bar\nu\gamma}^X$ is the $X$ contribution to the cross section
including $X$-SM interference terms.

With the much larger energies in this process than in the preceding low-energy cases,
it can offer access to the $m_X^{}$ dependence of the constraints on the $X$
couplings for larger values of $m_X^{}$ than the latter could.
In addition, the (anti)neutrinos now being only in the final state implies that all their
flavors can turn up.
Thus, \,$e^+e^-\to\nu\bar\nu\gamma$\, involves all the $X$ couplings to them, including
the one to $\nu_\tau^{}$, via $L_{\tau\tau}$ and $R_{\tau\tau}$ which do not
participate in the low-energy processes.

Since we are interested in applying the LEP data for \,$m_X^2<\hat s$,\,
the total-width $\Gamma_X$ needs to be taken into account.
However, in our model-independent analysis, its value is unknown, as we leave the $X$ couplings
to other SM particles unspecified and also it may have a component arising from decay
channels into final states comprising other nonstandard particles.
Consequently, we will assume particular values of $\Gamma_X$ for illustration.

With $\Gamma_X$ specified, it is important to ensure that the extracted ranges of $L_{ij}^{}$
and $R_{ij}^{}$ satisfy the requirement that the sum of $\Gamma_{X\to e^+e^-}$ and the rates
of all \,$X\to\nu_i\bar\nu_j$\, modes not exceed~$\Gamma_X$.
It is straightforward to realize that this amounts to demanding\footnote{\baselineskip=14pt%
With $\varepsilon$ and $\upsilon$ representing, respectively, the two factors on the left-hand
side of Eq.\,(\ref{GXlimit}), we can always write
\,$4\varepsilon\upsilon\le(\varepsilon+\upsilon)^2\le\Gamma_X^2$.}
\begin{eqnarray} \label{GXlimit}
\Gamma_{X\to e^+e^-}^{}\sum_{i,j=e,\mu,\tau}\Gamma_{X\to\nu_i^{}\bar\nu_j^{}}^{} \,\,\le\,\,
\mbox{$\frac{1}{4}$}\,\Gamma_X^2
\end{eqnarray}
with
\begin{eqnarray}
\Gamma_{X\to\nu_i^{}\bar\nu_j^{}}^{}\Gamma_{X\to e^+e^-}^{} \,\,=\,\,
\frac{\sqrt{m_X^2-4m_e^2}}{576\pi^2\,m_X^{}} \Bigl[
\Bigl(\bigl|L_{ij}^{}\bigr|^2+\bigl|R_{ij}^{}\bigr|^2\Bigr)\bigl(m_X^2-m_e^2\bigr) \,+\,
6\,L_{ij\,}^{}R_{ji\,}^{}m_e^2 \Bigr] ~.
\end{eqnarray}
For consistency, the value of $\Gamma_X$ picked in each instance needs to be sufficiently
large so that the constraints on $L_{ij}^{}$ and $R_{ij}^{}$ resulting from the application of
this condition are never stricter than the corresponding constraints imposed by
the \,$e^+e^-\to\nu\bar\nu\gamma$\, data.

\begin{figure}[b]
\includegraphics[width=143pt]{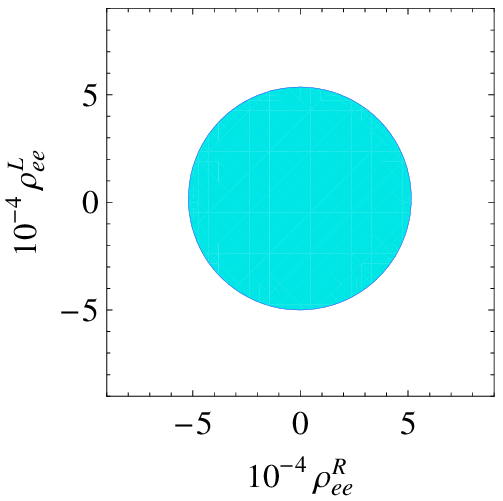} \, \, \,
\includegraphics[width=151pt]{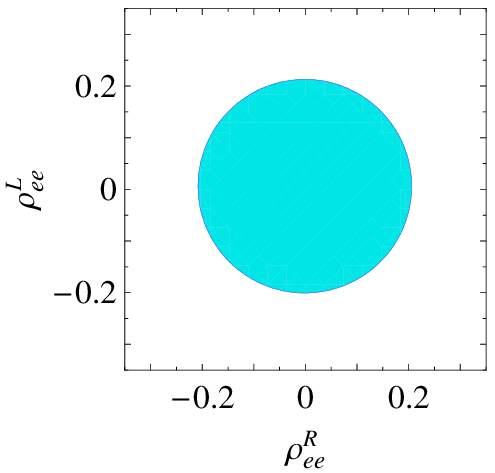} \, \, \,
\includegraphics[width=143pt]{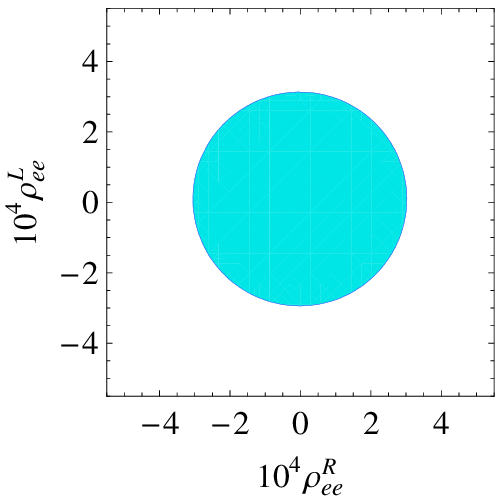}\vspace{-1ex}\\
\includegraphics[width=143pt]{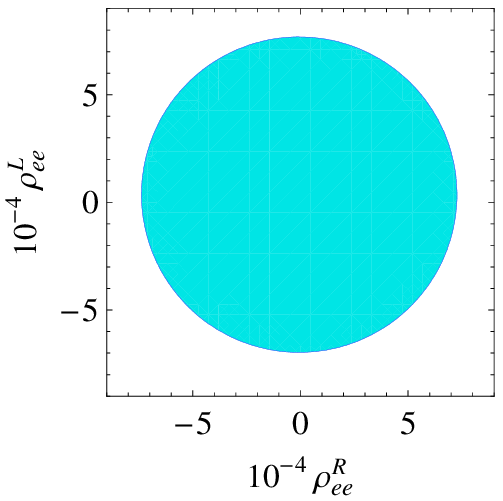} \, \, \,
\includegraphics[width=151pt]{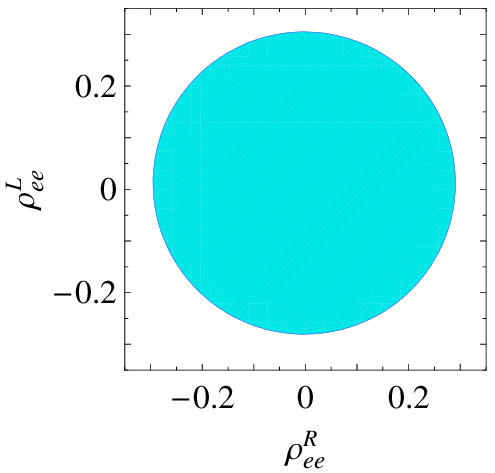} \, \, \,
\includegraphics[width=143pt]{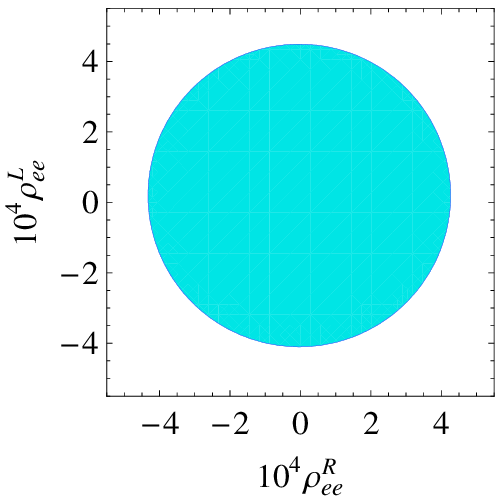}\vspace{-1ex}
\caption{Top plots: values of $\rho_{ee}^{L}$ and $\rho_{ee}^{R}$ (blue areas)
allowed by LEP data on~\,$e^+e^-\to\nu\bar\nu\gamma$\, and
the $\Gamma_X$ requirement in Eq.\,(\ref{GXlimit}) for,
from left to right,~\,$m_X^{}=0.01,5,100$~GeV\, and \,$\Gamma_X^{}=0.05,25,300$~keV,\,
respectively, in the limit that all the other $\rho_{ij}^{L,R}$ vanish.
Bottom plots: same as top ones, but for $\Gamma_X^{}$ being twice as
large.\label{ee2nngplots}}
\end{figure}

Based on our numerical exploration, we observe in general that for a fixed $m_X^{}$
the smaller the appropriately chosen value of $\Gamma_X$ is, the stronger the bounds from
\,$e^+e^-\to\nu\bar\nu\gamma$\, on the $X$ couplings, in accord with the expectation that
the partial rates which make up~$\Gamma_X$ rise and fall with the couplings.
This is illustrated with the blue (shaded) areas in Fig.\,\ref{ee2nngplots} for some values
of $m_X^{}$ and different $\Gamma_X^{}$ choices under the assumption that only
$\rho_{ee}^{L,R}$ are nonzero.
For these examples, from the top (bottom) plots we extract
\,$|L_{ee}|_{\rm max}^{1/2}\simeq|R_{ee}|_{\rm max}^{1/2}\simeq0.013,0.013,0.010$
$(0.016,0.016,0.012)$\, corresponding to~\,$m_X^{}=0.01,5,100$~GeV,\, respectively.
Assuming instead that only $\rho_{\mu\mu}^{L,R}$ or $\rho_{\tau\tau}^{L,R}$ are present,
we obtain results similar to those for $\rho_{ee}^{L,R}$, the difference being due to
small interference in the $ee$ case between the $X$- and $W$-mediated contributions.
These graphs also indicate that as $m_X^{}$ increases the constraints on
$\rho_{ee}^{L,R}$, $\rho_{\mu\mu}^{L,R}$, or $\rho_{\tau\tau}^{L,R}$ tend to get
stronger provided that $\Gamma_X^{}/m_X^{}$ does not change appreciably.

Moreover, comparing the $\rho_{ee,\mu\mu}^{L,R}$ plots with~Figs.~\ref{ll+reactors}(b)
and~\ref{charmplot}, respectively, we notice that for $m_X^{}$ values of a few GeV or
higher the areas permitted by the (anti)neutrino-electron scattering data can
significantly shrink to those around the origin after the inclusion of
the \,$e^+e^-\to\nu\bar\nu\gamma$\, constraints,
depending on~$\Gamma_X^{}$, and increasingly so as $m_X^{}$ goes up.
Thus, the two sets of data offer complementary restrictions on these $\rho$ parameters,
with the former (latter) yielding stronger constraints for lower (higher) masses.
Such complementarity is also visible in the limits on $L_{ee}$ and~$R_{ee}$, or their
combinations, in the special cases depicted in Fig.\,\ref{coupling-mass}, where the dotted
curves represent the limits from \,$e^+e^-\to\nu\bar\nu\gamma$.\,
These dotted curves correspond to the $\Gamma_X^{}$ choices, as in the top plots
of~Fig.\,\ref{ee2nngplots}, which are roughly the smallest ones satisfying the consistency
requirement mentioned after~Eq.\,(\ref{GXlimit}).

Before moving on, we note that in the family-universal limit,
\,$\rho_{ee}^{\sf C}=\rho_{\mu\mu}^{\sf C}=\rho_{\tau\tau}^{\sf C}$,\,  of the neutrino sector,
the allowed ranges of these ratios are somewhat smaller than in the family-nonuniversal case
due to the decrease in the number of free parameters.
In a specific model, the reduction of the $X$-coupling ranges is also generally expected to
happen because the model parameters are related to each other and subject to various
data~(see, {\it e.g.}, Refs.\,\cite{Boehm:2004uq,Fayet:2007ua}).

Another situation in which the constraints on the $X$ couplings can be stronger is when $X$
mixes with the SM gauge bosons.
In such a case, the mixing usually leads to a substantial increase in the number of experimental
observables that need to be taken into account with the mixing angle being the only additional
free parameter, and as a consequence the $X$ couplings become more
restrained.\footnote{\baselineskip=13pt%
This can occur in scenarios involving an extra U(1) gauge boson, such as considered
in~Refs.\,\cite{Williams:2011qb,Chiang:2011cv}.}
Thus, the numerical results of our analysis correspond to those in the limit that the mixing
is negligible.

\section{Constraints on separate flavor-conserving couplings\label{separate}}

There are observables that can yield bounds on the $X$ couplings to
the electron, $g_{Le,Re}^{}$, and neutrinos, $g_{\nu_i\nu_i}$, separately.
The resulting limits will then complement the limits on the products of
the couplings determined in the earlier sections.
To evaluate the most important constraints, we look at those from $e^+e^-$ scattering data
at the $Z$-pole and the measured anomalous magnetic moment of the electron, as well as
searches for a nonstandard spin-1 particle at fixed-target and beam-dump experiments.

The $Z$-pole observables with sensitivity to  $g_{Le,Re}^{}$ are the rate
\,$\Gamma_{Z\to e^+e^-}\simeq
\overline{|{\cal M}_{Z\to\bar e e}|^2}/\bigl(16\pi\hspace{1pt}m_Z^{}\bigr)$\,
and the parameter \,$A_e=\bigl(|L_e'|^2-|R_e'|^2\bigr)/\bigl(|L_e'|^2+|R_e'|^2\bigr)$\,
associated with the forward-backward asymmetry, which follow from the amplitude
\,${\cal M}_{Z\to\bar e e}=\bar e\!\not{\!\varepsilon}_Z^{}\bigl(L_e'P_L+R_e'P_R\bigr)e$.\,
The presence of $X$ causes modifications to the $Z e^+e^-$ vertex and
electron self-energy diagrams at the one-loop level.
Calculating the $X$ contributions and combining them with the SM ones, we have
\begin{eqnarray} \label{LeRe}
{\sf C}_e' &=& \frac{g\,\bar g_{\sf C}^{}}{c_{\rm w}^{}}
\bigl( 1+{\cal F}(\delta)\, g_{{\sf C}e}^2 \bigr) ~, \hspace{5ex}
{\sf C} \,\,=\,\, L,R ~, \hspace{5ex} \delta \,\,=\,\, \frac{m_X^2}{m_Z^2} ~,
\nonumber \\
{\cal F}(\delta) &=& \frac{1}{16\pi^2} \biggl\{
-\frac{7}{2} - 2\delta - (3+2\delta)\ln\delta -
2(1+\delta)^2\biggl[\ln\delta\;\ln\frac{\delta}{1+\delta}
+ {\rm Li}_2^{}\biggl(-\frac{1}{\delta}\biggr)\biggr]
\nonumber \\ && \hspace*{7ex} -\;
i\pi\, \biggl[ 3+2\delta+ 2(1+\delta)^2\,\ln\frac{\delta}{1+\delta} \biggr] \biggr\} ~,
\end{eqnarray}
where $\bar g_{L,R}^{}$ are defined in Eq.\,(\ref{Lsm}) and Li$_2$ is the dilogarithm.
The expression for the real part of ${\cal F}$ has been derived previously~\cite{Carone:1994aa}.
To probe $g_{\nu_i\nu_i}$, the relevant observable is \,$\Gamma_{Z\to\nu\bar\nu}$,\, which
dominates $\Gamma_{Z\to\rm invisible}$ and comes from an amplitude analogous to that in
\,$Z\to e^+e^-$,\, but without the right-handed coupling.
Since the neutrinos are not observed,
\begin{eqnarray} \label{GZ2nn}
\Gamma_{Z\to\nu\bar\nu}^{} \,\,=\,\,
\mbox{\footnotesize$\displaystyle\sum_{\scriptstyle i=e,\mu,\tau}$}\,
\Gamma_{Z\to\nu_i\bar\nu_i}^{}
\,\,=\,\, \frac{g^2m_Z^{}}{96\pi\,c_{\rm w}^2}\,
\mbox{\footnotesize$\displaystyle\sum_{\scriptstyle i=e,\mu,\tau}$}\,
\bigl|1+{\cal F}(\delta)\,g_{\nu_i\nu_i}^2\bigr|^2 ~.
\end{eqnarray}

\begin{figure}[b]
\includegraphics[width=85mm]{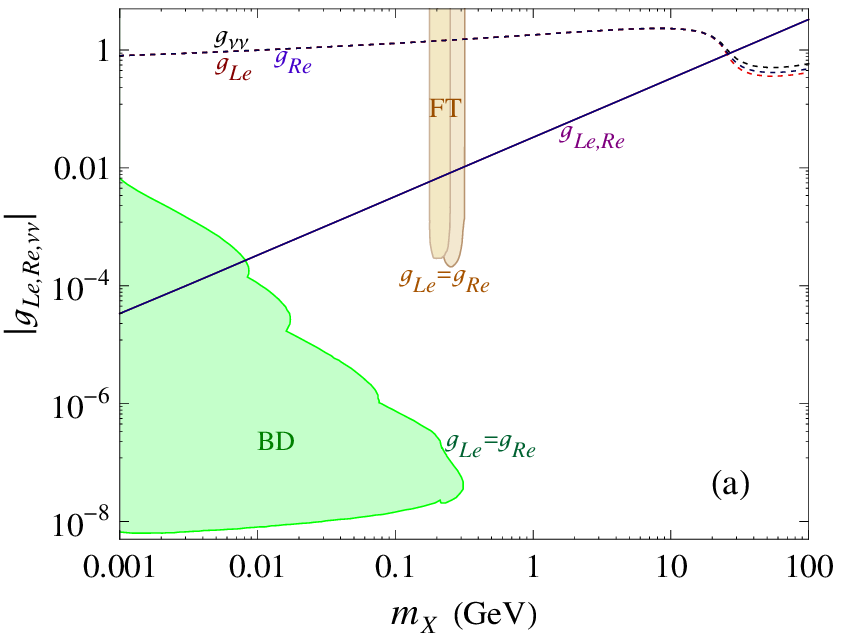} ~
\includegraphics[width=85mm]{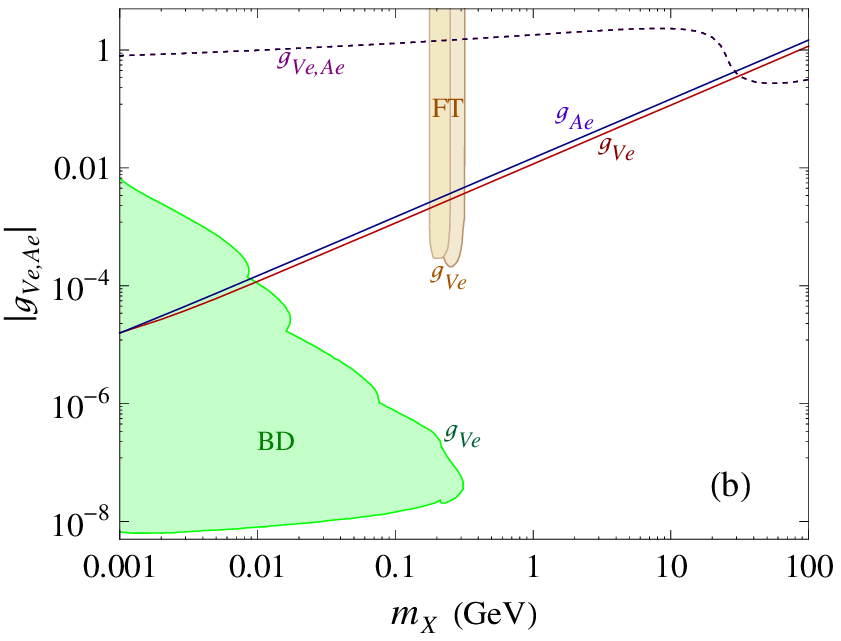} \vspace{-1ex}
\caption{The dotted curves describe the upper limits on (a)~$|g_{Le,Re,\nu_i\nu_i}^{}|$ and
(b)~$|g_{Ve,Ae}^{}|$ from $Z$-pole data under the assumption that only one of the couplings
is nonzero in each case, while the solid curves describe
the corresponding limits from the measured \,$g_e^{}-2$.\,
The light-orange and light-green regions contain values of
(a)~$g_{Le}^{}=g_{Re}^{}$ and (b)~$g_{Ve\,}^{}$ disallowed by electron-nucleus
fixed-target scattering (FT) and electron beam-dump (BD) experiments
under the assumption that \,$g_{Ae}^{}=0$\, and the branching
ratio~\,${\cal B}(X\to e^+e^-)=1$.\label{coupling-mass-2}}
\end{figure}

Now, the SM predicts that~\cite{erler-langacker} \,$\Gamma_{Z\to e^+e^-}^{\rm sm}=84.01\pm 0.07\;$MeV,\,
\,$A_e^{\rm sm}=0.1475\pm0.0010$,\, and \,$\Gamma_{Z\to\rm invisible}^{\rm sm}=501.69\pm0.06\;$MeV,\,
whereas experiments yield~\cite{pdg}
\,$\Gamma_{Z\to e^+e^-}^{\rm exp}=83.91\pm 0.12\;$MeV,\, \,$A_e^{\rm exp}=0.1515\pm0.0019$,\,
and \,$\Gamma_{Z\to\rm invisible}^{\rm exp}=499.0\pm1.5\;$MeV.\,
Accordingly, to restrain the $X$ couplings we can require them to satisfy the 90\%\,CL
ranges\footnote{\baselineskip=14pt%
We have taken the lower (upper) bound of $A_e$ $(\Gamma_{Z\to\nu\bar\nu})$ to be its SM
lower (upper) value because $A_e^{\rm exp}$ is above $A_e^{\rm sm}$
$\bigl(\Gamma_{Z\to\rm invisible}^{\rm exp}$ is below $\Gamma_{Z\to\rm invisible}^{\rm sm}\bigr)$
by {\small$\sim$\,}2 sigmas.}
\,$83.71{\rm\,MeV}\le\Gamma_{Z\to e^+e^-}\le84.11$\,MeV,\, \,$0.1459\le A_e\le0.1546$,\,
and~\,$497{\rm\,MeV}\le\Gamma_{Z\to\nu\bar\nu}\le502$\,MeV.\,
In extracting the couplings from these $Z$-pole measurements, for the SM parts we
employ the tree-level formulas along with the effective values \,$g_{\rm eff}^{}=0.6517$\, and
\,$s_{\rm w,eff}^2=0.23146$\, which lead to the $\Gamma_{Z\to e^+e^-}^{\rm sm}$ and
$A_e^{\rm sm}$ numbers above within their errors and
\,$\Gamma_{Z\to\nu\bar\nu}^{\rm sm}=501.26{\rm\,MeV}<\Gamma_{Z\to\rm invisible}^{\rm sm}$\,
in accord with expectation.
We show the results in Fig.\,\ref{coupling-mass-2} for $g_{Le,Re}^{}$, $g_{\nu_i\nu_i}^{}$, and
the combinations
\,$g_{Ve,Ae}^{}=\frac{1}{2}\bigl(g_{Le}^{}\pm g_{Re}^{}\bigr)$\, for the special cases in which
only one of them is nonzero (the dotted curves).
In Fig.\,\ref{coupling-mass-2}(b) the $g_{Ve,Ae}^{}$ (dotted) curves coincide, which can be
understood from the form of ${\sf C}_e$ in~Eq.\,(\ref{LeRe}).

The anomalous magnetic moment of the electron, \,$a_e^{}=g_e^{}/2-1$,\, has been measured
very precisely and therefore provides additional important constraints on $g_{Le,Re}^{}$
or~$g_{Ve,Ae}^{}$.
In terms of the latter, the $X$ contribution is~\cite{Leveille:1977rc}
\begin{eqnarray} \label{aeX}
a_e^X \,\,=\,\,
\frac{m^2_e}{4\pi^2m^2_X}\bigl(g_{V e}^2\,f_V^{}(r)+g_{A e}^2\,f_A^{}(r)\bigr) ~,
\end{eqnarray}
where \,$r=m^2_e/m^2_X$,\,
\begin{eqnarray}
f_V^{}(r) \;=\; \int^1_0 dx\; \frac{x^2-x^3}{1-x +r x^2} ~, \hspace{5ex}
f_A^{}(r) \;=\; \int^1_0 dx\; \frac{-4 x+5 x^2-x^3-2r x^3}{1-x +r x^2} ~.
\end{eqnarray}
The SM prediction for $a_e^{}$ is compatible with its latest measurement, the difference between
the two being \,$a_e^{\rm exp}-a_e^{\rm SM}=(-105\pm 82)\times10^{-14}$~\cite{Aoyama:2012wj}.
Consequently, for the $X$ contribution we can impose the 90\%\,CL
range~\,$-2.4\times10^{-12}\le a_e^X\le0.3\times10^{-12}$.\,
This translates into the limits (solid curves) graphed in Fig.\,\ref{coupling-mass-2}.
Evidently, these results are stricter (weaker) than those in the previous paragraph for
$m_X^{}$ below (above)~{\small$\sim$\,}30\,GeV.

There are also restrictions on $g_{Ve}^{}$ from the recent electron-nucleus fixed-target
scattering experiments by the A1 and APEX Collaborations~\cite{Merkel:2011ze},
but so far only for \,$m_X^{}=175$-300~MeV.\,
However, the strictness of the constraints depends on the assumed interactions of $X$ with
other particles.
Since the A1 and APEX analyses presupposed that $X$ coupled mainly to the electromagnetic
current (and hence had negligible axial-vector couplings to fermions),
following Ref.\,\cite{Bjorken:2009mm}, their limits on $g_{Ve}^{}$ apply most strongly
to the case in which the channel \,$X\to e^+e^-$\, highly dominates the $X$ decay.
In the limiting case that the branching ratio \,${\cal B}(X\to e^+e^-)=1$,
the excluded zone is displayed as the light-orange patch in
Fig.\,\ref{coupling-mass-2}(b)\footnote{\baselineskip=14pt%
Since the A1 and APEX bounds presupposed that $X$ coupled only to the electromagnetic
current~\cite{Merkel:2011ze,Bjorken:2009mm}, for the mass range
\,$2m_\mu^{}<m_X^{}\le300$\,MeV,\, before drawing the light-orange patch, we have lowered them
by a~factor of  \,${\cal S}=
\bigl[1+\Gamma_{X\to\mu^+\mu^-}\bigl(1+R(m_X)\bigr)/\Gamma_{X\to e^+e^-}\bigr]{}^{1/2}$\,
in order to account for the opening of the \,$\mu^+\mu^-$ and
$\pi^+\pi^-$\, decay channels of $X$.
Here $R$ is the energy-dependent ratio
\,$\sigma(e^+e^-\to{\rm hadrons})/\sigma(e^+e^-\to\mu^+\mu^-)$\, available
from~Ref.\,\cite{pdg} and the rate~\cite{Andreas:2012mt}
\,$\Gamma_{X\to l^+ l^-}=g_V^2m_X^{}\bigl(1+2 m_l^2/m_X^2\bigr)
\bigl(1-4 m_l^2/m_X^2\bigr){}^{1/2}/(12\pi)$\,
for \,$l=e,\mu$\, contains the vector coupling bound $g_V^{}$ supplied by A1 and
APEX~\cite{Merkel:2011ze}.\smallskip}
and the corresponding one in Fig.\,\ref{coupling-mass-2}(a), where for the latter
we have employed the fact that a limit on $g_{Ve}^{}$ in the absence of
\,$g_{Ae}^{}=\frac{1}{2}\bigl(g_{Le}^{}-g_{Re}^{}\bigr)$,\, as assumed by A1 and APEX in
interpreting their measurements~\cite{Merkel:2011ze}, implies a limit on \,$g_{Le}^{}=g_{Re}^{}$.
For \,$X\to e^+e^-$\, not being the dominant decay mode, \,${\cal B}(X\to e^+e^-)<1$,\,
such as in a complete study considering the $X$ couplings to all fermions,
the restraints would be lessened and the light-orange regions
move upward by a factor of \,$[{\cal B}(X\to e^+e^-)]^{-1/2}$.

For $m_X^{}$ under a few hundred MeV, further constraints might be available from electron
beam-dump experiments, where $X$ could be produced by bremsstrahlung from an electron
scattering off a nuclear target and pass through a~shield before decaying into an $e^+e^-$
pair in front of the detector~\cite{Bjorken:2009mm,Andreas:2012mt,Freytsis:2009bh}.
The $X$ decay length $l_X$ would then need to be between the length of the target
plus the shield and the total distance from the target to the detector.
In view of the latest information on the relevant past experiments collected
in~Ref.\,\cite{Andreas:2012mt}, including their electron beam energies and shield lengths,
one might ask if their data could bound the $X$ couplings.
Since \,$l_X^{}=p_X^{}/\bigl(\Gamma_X^{}m_X^{}\bigr)$,\, with $p_X^{}$ being the laboratory
momentum of~$X$,
the answer would depend on the choice of $\Gamma_X$ or the assumed interactions
of $X$, as in the last paragraph.
For instance, if $\Gamma_X$ has the values chosen for the dotted curves
in~Fig.\,\ref{coupling-mass}, we determine that the electron beam-dump experiments listed
in~Ref.\,\cite{Andreas:2012mt} could not probe $X$ because it would decay well
inside their shields.
On the other hand, if $\Gamma_X$ is smaller and within the appropriate range, it would be
possible for $X$ to decay in the detection region depending on~$m_X^{}$.
In that case, we can look at the latest limits supplied in~Ref.\,\cite{Andreas:2012mt} whose
authors also assumed that $X$ effectively coupled only to the electromagnetic current and
thus had vanishing axial-vector couplings.
For~\,${\cal B}(X\to e^+e^-)=1$,\, their results translate into
the disfavored parameter space represented by the light-green portions
of~Fig.\,\ref{coupling-mass-2},\footnote{\baselineskip=14pt%
For \,$m_X^{}>2m_\mu^{}$,\, we have again reduced the bounds from
Ref.\,\cite{Andreas:2012mt} by the corresponding factor $\cal S$.\vspace{-1ex}}
while for \,$X\to e^+e^-$\, not being dominant the restrictions would decrease and
the light-green zones would shift upward also by a~factor of~$[{\cal B}(X\to e^+e^-)]^{-1/2}$.

Some additional constraints may come from the measurements of \,$e^+e^-\to e^+e^-l^+l^-$\,
scattering for~\,$l=e,\mu,\tau$,\, but the available data, collected at energies around
the $Z$ resonance by the ALEPH Collaboration~\cite{Buskulic:1994gk}, are rather
limited compared to most of those discussed above.
Since the observed numbers of events for the different leptonic final-states
were consistent within up to about~40\% with the SM
expectations~\cite{Buskulic:1994gk}, to extract the constraints on the $X$ couplings to
the electron we may require that their effects be less than 20\% of the SM contributions.
To estimate the cross section including the $X$ contribution, we employ the CalcHEP
package~\cite{Pukhov:1999gg}.
The resulting constraints depend again on the choices of \,${\cal B}(X\to e^+e^-)$.
If \,${\cal B}(X\to e^+e^-)=1$\, and only one of the $X$ couplings is nonzero at a time,
we get \,$|g_{Le,Re}^{}|$\, below roughly \,$0.2,\,0.08,\,0.5$\, for
\,$m_X^{}=0.01,1,100$~GeV,\, respectively.
The bounds would be weaker if~\,${\cal B}(X\to e^+e^-)<1$.\,
Hence they generally are not more stringent than the strictest bounds exhibited
in~Fig.\,\ref{coupling-mass-2}.

The results presented in Figs.~\ref{coupling-mass} and \ref{coupling-mass-2} illustrate how
the various measurements complement each other in probing the $X$ couplings.
If the electron and neutrino couplings of $X$ are not very dissimilar in size,
the (anti)neutrino-electron scattering and \,$e^+e^-\to\nu\bar\nu\gamma$\, data can be
expected to offer the strictest constraints.
If instead one of the electron and neutrino couplings is much larger than the other, then
the $Z$-pole data and measured $a_e^{}$ may provide the best constraints, depending on~$m_X^{}$.
For a sub-GeV mass, electron fixed-target and beam-dump experiments can yield the most stringent
test in the case of $X$ dominantly coupling to the electron.
Furthermore, if $m_X^{}$ is a few GeV or lower, future searches of \,$e^+e^-\to\gamma X$,\,
$X\to e^+e^-$\,
at high-luminosity colliders may offer competitive probes~\cite{Reece:2009un}.
We note that the case of the electron couplings being much larger than the neutrino couplings
could occur even if $X$ hails from an extra gauge sector, one example being the dark/hidden
photon, which is a~spin-1 mass-eigenstate associated with a new U(1) symmetry and coupling
predominantly to SM charged fermions through the electromagnetic
current~\cite{Bjorken:2009mm,Andreas:2012mt}.

Also, our results are applicable to some of the scenarios mentioned in Section \ref{intro}
which may explain certain experimental anomalies.
In particular, our bounds on the $X$ couplings already probe parts of the parameter space of
the spin-1 particles that may have been the factors behind the unexpected observations of
the 511-keV emission from our galactic bulge and of the positron excess in cosmic rays,
but the bounds do not yet test the validity of these scenarios.
In the former case, $X$ has mass of ${\cal O}(1$\,MeV) and may be detectable with future
neutrino telescopes if
\,$g_{\nu_i\nu_i}^{}\mbox{\footnotesize\,$\gtrsim$\,}10^{-5}$\,~\cite{Hooper:2007jr}, which
easily satisfies the $g_{\nu_i\nu_i}^{}$ limit in Fig.\,\ref{coupling-mass-2}(a) and implies
that $g_{Le,Re}^{}$ need to be sufficiently small to evade the restrictions depicted
in~Fig.\,\ref{coupling-mass}(a).
In the latter scenario, $m_X^{}$ is of order a few GeV and the corresponding limits given in
Figs.$\;$\ref{coupling-mass} and~\ref{coupling-mass-2} indicate the level of constraints on
$g_{Le,Re,\nu_i\nu_i}^{}$ from current data, but future searches in the BESIII experiment
could be sensitive to $g_{Le,Re}^{}$ and $g_{\nu_i\nu_i}^{}$  as small as $10^{-4}$
and $10^{-5}$, respectively \cite{Foot:1994vd}.

%%%%%%%%%%%%%%%%%%%%%%%%%%%%%%%%%%%%%%%%%%%%%%%%%%
\section{Flavor-changing couplings\label{fcc}}
%%%%%%%%%%%%%%%%%%%%%%%%%%%%%%%%%%%%%%%%%%%%%%%%%%

Finally, we explore the bounds on the flavor-changing parameters, $L_{ij}$ and $R_{ij}$ for
\,$i\neq j$,\, as well as the corresponding $\rho$ parameters,
from the same experimental inputs as considered in~Sections~\ref{nue}-\ref{ee2nng}.
To do so, we assume that only one of these $(L,R)$ pairs is nonzero at a time and turn off
all the flavor-conserving ones, \,$L_{ii}=R_{ii}=0$.\,
Furthermore, here we slightly modify the definition of $L_{ij}$ to
\,$L_{ij}=|g_{\nu_i\nu_j}|\,g_{Le}^{}=L_{ji}$\, to make it real, and similarly with~$R_{ij}$.

\begin{figure}[b]
\includegraphics[width=160pt]{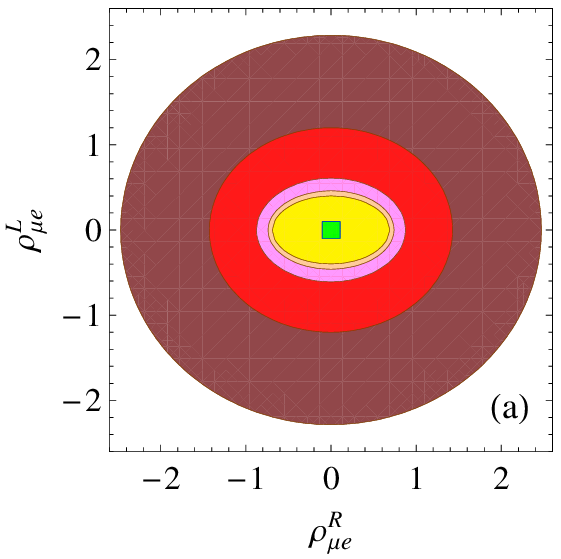}
\includegraphics[width=165pt]{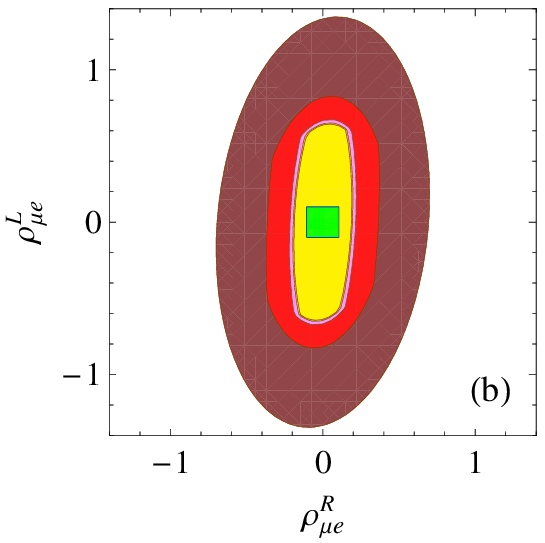} \!\!\!\!
\includegraphics[width=165pt]{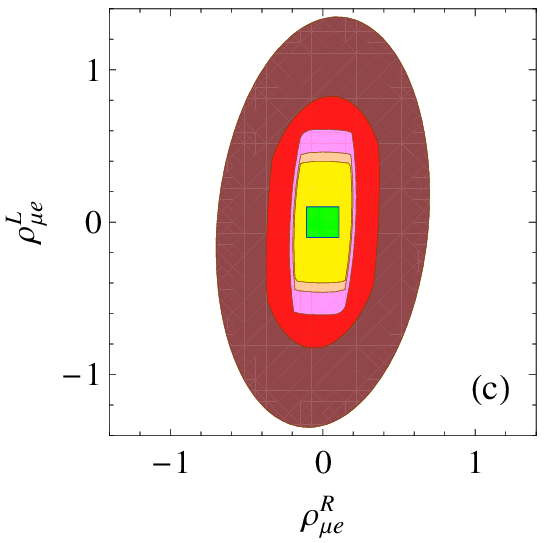}\vspace{-1ex}
\caption{Values of $\rho_{\mu e}^{L}$ and $\rho_{\mu e}^{R}$ allowed by
(a)~\,$\nu_{e,\mu}^{}e\to\nu e$\, data, (b)~\,$\bar\nu_{e,\mu}^{}e\to\bar\nu e$\, data, and
(c)~all of them for \,$m_X^{}=1$\,MeV\,(brown, darkest colored),
2\,MeV\,(red), 5\,MeV\,(magenta), 10\,MeV\,(orange), 50\,MeV\,(yellow, most lightly-shaded),
and \,$m_X^{}\mbox{\small\,$\gtrsim$\,}1$\,GeV (smallest green areas)\,
in the limit that all the other $\rho_{ij}^{L,R}$ vanish.\label{fc-ne2ne-plots}}
\end{figure}

We display in Fig.\,\ref{fc-ne2ne-plots} the values of \,$\rho_{\mu e}^{L,R}=\rho_{e\mu}^{L,R}$\,
permitted by the $\nu_e^{}e$ and $\bar\nu_e^{}e$ scattering data and,
if~\,$m_X^{}\mbox{\small\,$\gtrsim$\,}1$\,GeV,\, by the $\nu_\mu^{}e$ and~$\bar\nu_\mu^{}e$
scattering data.
As Fig.\,\ref{fc-ne2ne-plots}(c) indicates, we find that the maximal values
$\bigl|\rho_{e\mu}^{L(R)}\bigr|{}_{\rm max}^{}$ vary from 1.3 to 0.10 (0.70 to 0.11)
as $m_X^{}$ rises from 1\,MeV to 1\,GeV,\, and accordingly
$|L_{e\mu}|_{\rm max}^{1/2}$ $\bigl(|R_{e\mu}|_{\rm max}^{1/2}\bigr)$ varies
from  \,$7\,(5)\times10^{-6}$\, to~\,0.002\,(0.002)\, in this mass range.
If \,$m_X^{}=5\,(100)$~GeV\, instead, one would find
\,$|L_{e\mu}|_{\rm max}^{1/2}\simeq|R_{e\mu}|_{\rm max}^{1/2}${\small\,$\sim$\,}0.004\,(0.02)\,
from the smallest (green) area.
For the $e\tau$ parameters, the allowed regions are the same as those for $e\mu$ subject to the
$\nu_e^{}e$ and $\bar\nu_e^{}e$ data.
Thus, $\bigl|\rho_{e\tau}^{L(R)}\bigr|{}_{\rm max}^{}$ changes from 1.3 to 0.40 (0.70 to 0.19)
for \,$m_X^{}=1$-50~MeV,\, as Fig.\,\ref{fc-ne2ne-plots}(c) shows, and the corresponding
numbers for \,$m_X^{}\mbox{\small\;$\gtrsim$\;}40$\,MeV\, are practically the same as those
for \,$m_X^{}=50$\,MeV.\,
The limits on~$\rho_{\tau\mu,\mu\tau}^{L,R}$ come from $\nu_\mu^{}e$ and~$\bar\nu_\mu^{}e$
scattering, and so \,$\bigl|\rho_{\mu\tau}^{L,R}\bigr|{}_{\rm max}^{}\simeq0.1$\,
for~\,$m_X^{}\mbox{\small\,$\gtrsim$\,}1$\,GeV.\,
It is evident from these examples that the patterns of low-$m_X^{}$ dependence seen in
the flavor-conserving cases roughly turn up again here.

\begin{figure}[t]
\includegraphics[width=150pt]{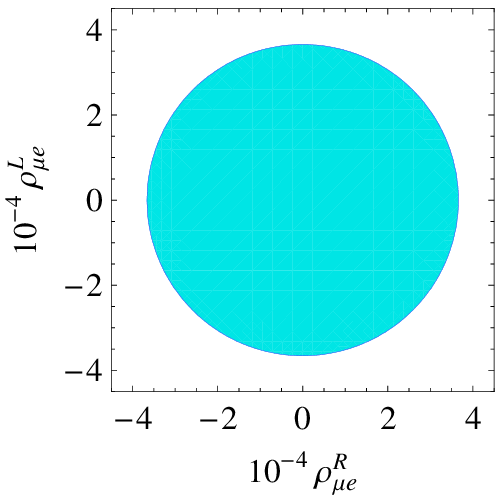} \, \,
\includegraphics[width=157pt]{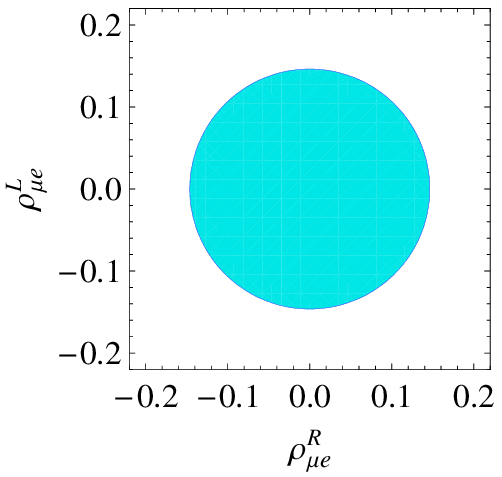} \, \,
\includegraphics[width=150pt]{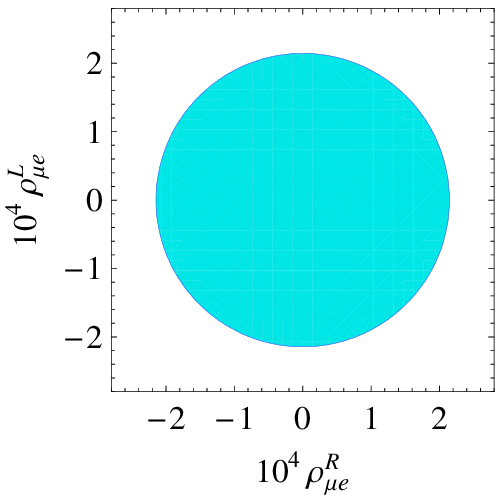}\vspace{-1ex}
\caption{Values of $\rho_{\mu e}^{L,R}=\rho_{e\mu}^{L,R}$ allowed by
\,$e^+e^-\to\nu\bar\nu\gamma$\, data and $\Gamma_X$ requirement in Eq.\,(\ref{GXlimit})
for, from left to right,~\,$m_X^{}=0.01,5,100$~GeV\, and \,$\Gamma_X^{}=0.05,25,300$~keV,\,
respectively, if all the other $\rho_{ij}^{L,R}$ vanish.\label{fc-ee2nng-plots}}
\end{figure}

In Fig.\,\ref{fc-ee2nng-plots} we present the ranges of \,$\rho_{\mu e}^{L,R}=\rho_{e\mu}^{L,R}$\,
satisfying the \,$e^+e^-\to\nu\bar\nu\gamma$\, restrictions (blue shaded areas) for
the same $m_X^{}$ and $\Gamma_X$ choices as in the top plots in~Fig.\,\ref{ee2nngplots}.
In drawing Fig.\,\ref{fc-ee2nng-plots}, we have taken into account the fact that
\,$\rho_{\mu e}^{L,R}$\, and \,$\rho_{e\mu}^{L,R}$\, contribute to the cross section via a pair
of charge-conjugate final states.
For these instances, the boundaries of the blue regions imply
\,$|L_{\mu e}|_{\rm max}^{1/2}=|R_{\mu e}|_{\rm max}^{1/2}\simeq0.011,0.011,0.008$\,
for~\,$m_X^{}=0.01,5,100$~GeV,\, respectively.
These results are the same as those for the corresponding $e\tau$ and $\mu\tau$ parameters.

The examples in the last two paragraphs demonstrate as in Section~\ref{ee2nng} that
the \,$e^+e^-\to\nu\bar\nu\gamma$\, constraints can be stronger than those from
$\nu e$ and $\bar\nu e$ scattering data, depending on~$m_X^{}$ and~$\Gamma_X$, especially
for $m_X^{}$ above~1\,GeV.
Hence again the two sets of measurements are complementary to each other in bounding
the $X$ leptonic couplings.

Lastly, we note that these flavor-changing parameters can induce at the one-loop
level flavor-violating $X$-mediated transitions involving charged leptons,
such as \,$\mu\to 3e$\, and \,$\tau\to3e$.\,
The loop consists of internal $W$ and flavor-changing neutrino lines, and $X$ is attached to
the neutrino line.
Although the experimental upper limits of their branching ratios are very stringent, whether
or not they could yield strict bounds on the flavor-changing $X$ couplings depends on whether
the complete model possesses a Glashow-Iliopoulos-Maiani-like mechanism in the lepton sector.
If it does, the branching ratios will be very suppressed by the neutrino
masses, as in Eq.\,(\ref{l2lll}), allowing the couplings to evade the experimental limits.
However, if such a mechanism is absent, the couplings will be highly suppressed compared to
the results found above.

%%%%%%%%%%%%%%%%%%%%%%%%%%%%%%%%%%%%%%%%%%%%%%%%%%
\section{Conclusions\label{sec:summary}}
%%%%%%%%%%%%%%%%%%%%%%%%%%%%%%%%%%%%%%%%%%%%%%%%%%

There has been some regained interest in new light spin-1 particles, with mass in the regime
of several GeV to sub-GeV, in the hope of explaining a number of experimental anomalies and
unexpected astronomical observations.
We have explored in this work the constraints on the neutrino and electron interactions of
such a particle from several sets of lepton scattering data.
The new boson, $X$, is assumed to be electrically and color neutral, without mixing with
the standard-model gauge bosons, and its couplings with the leptons are taken to be
sufficiently general for a~model-independent approach.
Our analysis starts with the case of only flavor-conserving couplings.
We utilize the \,$\nu_e^{}e^-\to\nu e^-$\, scattering data acquired in the E225 experiment
at LAMPF and the LSND experiment, in tandem with the \,$\bar\nu_e^{}e^-\to\bar\nu e^-$\,
data obtained from several experiments at nuclear power plants, to place bounds on the chiral
couplings of $X$ to the electron neutrino and electron, via the $\rho_{ee}^{L,R}$ parameters
defined in the main text.
We illustrate how, for a relatively light $X$, its mass may substantially affect
the determination of its couplings to these leptons.
Subsequently, constraints on their muon-neutrino counterparts are imposed by means of
the CHARM-II data on \,$\nu_\mu^{}e^-\to\nu e^-$\, and \,$\bar\nu_\mu^{}e^-\to\bar\nu e^-$\,
scattering.
The LEP measurements of \,$e^+e^-\to\nu\bar\nu\gamma$\, are then employed to derive
a~complementary set of experimental bounds on $\rho_{ee,\mu\mu}^{L,R}$, plus the only
restraints on~$\rho_{\tau\tau}^{L,R}$, the result of which shows significant dependence on
the mass and decay width of the $X$ boson.
As an important supplement, we also evaluate constraints on the respective flavor-conserving $X$
couplings to the electron and neutrinos from $Z$-pole data, the measured $g_e^{}-2$, and
searches at fixed-target and beam-dump experiments.
Finally, we apply the same inputs from the (anti)neutrino-electron
and \,$e^+e^-\to\nu\bar\nu\gamma$\, scattering experiments to the case where only one pair of
flavor-violating chiral couplings of $X$ is dominant to find their allowed ranges.
In summary, under our assumptions, the current experimental data restrict the couplings
within narrow regions consistent with zero over a~wide range of the new boson mass.

%%%%%%%%%%%%%%%%%%%%%%%%%%%%%%%%%%%%%%%%%%%%%%%%%%
\acknowledgments
%%%%%%%%%%%%%%%%%%%%%%%%%%%%%%%%%%%%%%%%%%%%%%%%%%

We would like to thank Sechul~Oh for conversations which led to this work.
We also thank Takaaki Nomura and Kei Yagyu for assistance with CalcHEP.
This work was supported in part by National Center for Theoretical Sciences,
the National Science Council of R.O.C. under Grants Nos.
NSC-100-2628-M-008-003-MY4, NSC-99-2112-M-008-003-MY3, and NSC-100-2811-M-008-036, and
the NCU Plan to Develop First-Class Universities and Top-Level Research Centers.

%%%%%%%%%%%%%%%%%%%%%%%%%%%%%%%%%%%%%%%%%%%%%%%%%%
\appendix
%%%%%%%%%%%%%%%%%%%%%%%%%%%%%%%%%%%%%%%%%%%%%%%%%%

%%%%%%%%%%%%%%%%%%%%%%%%%%%%%%%%%%%%%%%%%%%%%%%%%%
\section{Squared amplitudes and \,$\bm{e^+e^-\to\nu\bar\nu\gamma}$ data\label{formulas}}
%%%%%%%%%%%%%%%%%%%%%%%%%%%%%%%%%%%%%%%%%%%%%%%%%%

The tree-level contribution of the SM to the amplitude for
\,$\nu_i^{}e^-\to\nu_j^{}e^-$\, with \,$i=j=e$\, arises from $u$-channel $W$-mediated and
$t$-channel $Z$-mediated diagrams.  The $X$-mediated diagram contributes in the $t$-channel.
For \,$i=j=\mu$\, the $W$-mediated contribution is absent, while for \,$i\neq j$\,
only the $X$ contribution is present.
Neglecting the neutrino mass, averaging the absolute square of the amplitude over the initial
electron spins, the incident neutrino being left-handed, and
summing the amplitude over the final spins, we then arrive at for \,$i=e$ or $\mu$\,
\begin{eqnarray} \label{M2nuee}
\overline{\bigl|{\cal M}_{\nu_i^{}e\to\nu_i^{}e}\bigr|^2} \,\,=\,\,
\overline{\bigl|{\cal M}_{\nu_i^{}e\to\nu_i^{}e}^{\rm SM}\bigr|^2} \,+\,
\overline{\bigl|{\cal M}_{\nu_i^{}e\to\nu_i^{}e}\bigr|_X^2} ~,
\end{eqnarray}
\begin{eqnarray} \label{Msm2nuee}
\overline{\bigl|{\cal M}_{\nu_i^{}e\to\nu_i^{}e}^{\rm SM}\bigr|^2}  &=&
\frac{\omega\,g^4}{2\,{\cal U}_W^2} \Biggl[ \bigl(s-m_e^2\bigr)^{\!2} +
\frac{m_{e\,}^4 t}{m_W^2} + \frac{m_e^4\bigl(u-m_e^2\bigr){}^2}{4 m_W^4} \Biggr]
\nonumber \\ && +\;
\frac{\omega\,g^4}{c_{\rm w}^2\,{\cal U}_W^{}{\cal T}_Z^{}} \Biggl\{
\bar g_L^{} \Biggl[ \bigl(s-m_e^2\bigr)^{\!2} + \frac{m_{e\,}^4 t}{2 m_W^2} \Biggr] +
\bar g_{R\,}^{}m_e^2 \Biggl[ t + \frac{\bigl(u-m_e^2\bigr){}^2}{2 m_W^2} \Biggr] \Biggr\}
\nonumber \\ && +\;
\frac{g^4}{2c_{\rm w}^4\,{\cal T}_Z^2} \Bigl[ \bar g_L^2\,\bigl(s-m_e^2\bigr){}^2 +
\bar g_R^2\,\bigl(u-m_e^2\bigr){}^2 + 2\,\bar g_{L\,}^{}\bar g_{R\,}^{} m_{e\,}^2 t \Bigr] ~,
\end{eqnarray}
\begin{eqnarray} \label{MX2nuee}
\overline{\bigl|{\cal M}_{\nu_i^{}e\to\nu_i^{}e}\bigr|_X^2}  &=&
\frac{2\omega\,g^2}{{\cal U}_W^{}{\cal T}_X^{}} \Biggl\{
L_{ee}^{}    \Biggl[ \bigl(s-m_e^2\bigr)^{\!2}+\frac{m_{e\,}^4 t}{2 m_W^2} \Biggr] +
R_{ee}^{} m_e^2       \Biggl[ t+\frac{\bigl(u-m_e^2\bigr){}^2}{2 m_W^2} \Biggr] \Biggr\}
\nonumber \\ && +\;
\frac{2\,g^2}{c_{\rm w}^2\,{\cal T}_Z^{}{\cal T}_X^{}} \Bigl\{
\bar g_L^{} \Bigl[ L_{ii}^{}\bigl(s-m_e^2\bigr){}^2 + R_{ii\,}^{}m_{e\,}^2 t \Bigr] +
\bar g_R^{} \Bigl[ R_{ii}^{}\bigl(u-m_e^2\bigr){}^2 + L_{ii\,}^{}m_{e\,}^2 t \Bigr] \Bigr\}
\nonumber \\ && +\;
\frac{2}{{\cal T}_X^2}\Bigl[ L_{ii}^2\bigl(s-m_e^2\bigr){}^2 +
R_{ii}^2\bigl(u-m_e^2\bigr){}^2 + 2\,L_{ii}^{}R_{ii\,}^{}m_{e\,}^2 t \Bigr] ~,
\end{eqnarray}
\begin{eqnarray} & \displaystyle
s \,\,=\,\, (p_\nu^{}+p_e^{})^2 ~, \hspace{5ex} t \,\,=\,\, (p_e'-p_e^{})^2 ~, \hspace{5ex}
u \,\,=\,\, 2m_e^2-s-t ~, & \nonumber \\  & \displaystyle
{\cal T}_P^{} \,\,=\,\, t-m_P^2 ~, \hspace{5ex} {\cal U}_P^{} \,\,=\,\, u-m_P^2 ~, \hspace{5ex}
{\sf C}_{ii}^{} \,\,=\,\, g_{\nu_i^{}\nu_i^{}\,}^{}g_{{\sf C}e}^{} ~, \hspace{3ex}
{\sf C} \,\,=\,\, L,R ~, &
\end{eqnarray}
where \,$\omega=1\,(0)$\, if \,$i=e\,(\mu)$,\, the expression in Eq.\,(\ref{MX2nuee}) contains
$X$-SM interference terms plus a~purely $X$-induced part, and
$p_\nu^{}$ and $p_e^{}$ $\bigl(p_\nu'$ and $p_e'\bigr)$ are the four-momenta of the initial
(final) neutrino and electron, respectively.
For \,$j\neq i$
\begin{eqnarray} \label{M2nule}
\overline{\bigl|{\cal M}_{\nu_i^{}e\to\nu_j^{}e}\bigr|^2} \,\,=\,\,
\frac{2}{{\cal T}_X^2}\Bigl[ |L_{ji}^{}|^2\bigl(s-m_e^2\bigr){}^2 +
|R_{ji}^{}|^2\bigl(u-m_e^2\bigr){}^2 + 2\,L_{ij}^{}R_{ji\,}^{}m_e^2\,t \Bigr] \,, \hspace{4ex}
{\sf C}_{ij}^{} \,=\, g_{\nu_i^{}\nu_j^{}\,}^{}g_{{\sf C}e}^{} ~, ~~~~
\end{eqnarray}
where we have used~\,${\rm Re}\bigl(L_{ji}^*R_{ji}^{}\bigr)=L_{ij}R_{ji}$\,
following from \,$g_{\nu_j\nu_i}^*=g_{\nu_i\nu_j}^{}$.\,
In the laboratory frame where the initial electron is at rest,
\begin{eqnarray}
s \,\,=\,\, 2 E_\nu^{}m_e^{}+m_e^2 ~, \hspace{5ex} t \,\,=\,\, -2m_e^{}T ~,
\end{eqnarray}
where \,$E_\nu^{}$ is the energy of the incident neutrino and $T$ the kinetic energy of
the recoiling electron.
From the scattering kinematics, it is simple to show~\cite{Bahcall:1964zzb}
\begin{eqnarray} \label{Trange}
0 \,\,\le\,\, T \,\,\le\,\, \frac{2 E_\nu^2}{2E_\nu^{}+m_e^{}} ~.
\end{eqnarray}

For the \,$\bar\nu_i^{}e^-\to\bar\nu_j^{}e^-$\, scattering with \,$i=j=e$,
the amplitude receives contributions from SM $s$-channel $W$-mediated and $t$-channel
$Z$-mediated diagrams and a $t$-channel $X$-mediated diagram.
As in the preceding paragraph, for \,$i=j=\mu$\, the $W$-mediated diagram is absent,
whereas for \,$j\neq i$\, only the $X$ contribution is present.
It follows that from Eqs.~(\ref{M2nuee})-(\ref{MX2nuee}) and~(\ref{M2nule}) we can derive
the corresponding formulas for \,$\bar\nu_i^{}e^-\to\bar\nu_i^{}e^-$\, and
\,$\bar\nu_i^{}e\to\bar\nu_j^{}e$\, with \,$i\neq j$,\, respectively, by simply
interchanging $s$ and $u$, assuming~\,$s<m_W^2$.\,
From the $\bar\nu e$ counterpart of Eq.\,(\ref{Trange}), it is straightforward to
obtain the minimum energy of the incident antineutrino~\cite{Bahcall:1964zzb}
\begin{eqnarray} \label{Enumin}
2E_{\bar\nu}^{\rm min} \,\,=\,\, T+\sqrt{2m_e^{}T+T^2}
\end{eqnarray}
for a given $T$.

For the \,$e^+e^-\to\bar\nu\nu\gamma$\, scattering, the amplitude receives contributions from
five tree-level diagrams in the SM, three of which are mediated by the $W$ and two by the~$Z$,
and from two $X$-mediated diagrams similar to the $Z$ diagrams,
with the final-state photon being attached to the $e^\pm$ and $W$ lines.
The invariant kinematical variables can be chosen to be~\cite{Berends:1987zz}
\begin{eqnarray} &
\hat s \,\,=\,\, \bigl(p_{e^+}^{}+p_{e^-}^{}\bigr)^2 ~, ~~~~~~~
\hat t \,\,=\,\, \bigl(p_{e^+}^{}-p_{\bar\nu}^{}\bigr)^2 ~, ~~~~~~~
\hat u \,\,=\,\, \bigl(p_{e^+}^{}-p_\nu^{}\bigr)^2 ~, & \\ &
\hat s' \,\,=\,\, \bigl(p_{\bar\nu}^{}+p_\nu^{}\bigr)^2 ~, ~~~~~~~
\hat t' \,\,=\,\, \bigl(p_{e^-}^{}-p_\nu^{}\bigr)^2 ~, ~~~~~~~
\hat u' \,\,=\,\, \bigl(p_{e^-}^{}-p_{\bar\nu}^{}\bigr)^2 ~, & \\ &
\kappa_+^{} \,\,=\,\, 2 p_{e^+}^{}\cdot p_\gamma^{} ~, ~~~~~~~
\kappa_-^{} \,\,=\,\, 2 p_{e^-}^{}\cdot p_\gamma^{} ~, ~~~~~~~
\kappa_+' \,\,=\,\, 2 p_{\bar\nu}^{}\cdot p_\gamma^{} ~, ~~~~~~~
\kappa_-' \,\,=\,\, 2 p_\nu^{}\cdot p_\gamma^{} ~, & ~~~~~~~
\end{eqnarray}
where $p_{e^\pm}$ are the four-momenta of $e^\pm$, etc.
Averaging (summing) the absolute square of the amplitude over initial (final) spins and
including all neutrino flavors, one then finds the contributions to the cross section
in~Eq.\,(\ref{cs_ee2nng})
\begin{eqnarray} \label{M2ee2nng}
\overline{\bigl|{\cal M}_{e\bar e\to\nu_e\bar\nu_e\gamma}^{}\bigr|^2}  &=&
\frac{e^2 g^4}{2\kappa_-^{}\kappa_+^{}}\Biggl\{
\Biggl[\biggl|2{\cal G}_L^{ee}+\frac{1}{\cal W}\biggr|^2\hat u^2 +
\biggl|2{\cal G}_L^{ee}+\frac{1}{\cal W'}\biggr|^2\hat u^{\prime2} +
\bigl|2{\cal G}_R^{ee}\bigr|^2\bigl(\hat t^2+\hat t^{\prime2}\bigr) \Biggr]\hat s'
\nonumber \\ && \hspace*{9ex} -\;
{\rm Re}\Biggl[\biggl(2{\cal G}_L^{ee}+\frac{1}{\cal W}\biggr)^{\!\!*}\,
\frac{\kappa_-^{}\hat s'-\kappa_+'\hat t'+\kappa_-'\hat u'
+ 4i\,\epsilon_{\eta\rho\tau\omega}^{}\,
p_{e^-}^\eta p_\nu^\rho\,p_{\bar\nu}^\tau\,p_\gamma^\omega}{\cal W W'}\Biggr]u^2
\nonumber \\ && \hspace*{9ex} -\;
{\rm Re}\Biggl[\biggl(2{\cal G}_L^{ee}+\frac{1}{\cal W'}\biggr)^{\!\!*}\,
\frac{\kappa_+^{}\hat s'-\kappa_-'\hat t+\kappa_+'\hat u
+ 4i\,\epsilon_{\eta\rho\tau\omega}^{}\,
p_{e^+}^\eta p_\nu^\rho\,p_{\bar\nu}^\tau\,p_\gamma^\omega}{\cal W W'}
\Biggr]u^{\prime2}
\nonumber \\ && \hspace*{9ex} -\;
\frac{\kappa_-^{}\,\kappa_+'\,\hat t'\hat u^2+\kappa_+^{}\,\kappa_-'\,\hat t\,\hat u^{\prime2}}
{|{\cal W W'}|^2} \Biggr\} ~,
\end{eqnarray}
\begin{eqnarray} \label{M2ee2nng'}
\overline{\bigl|{\cal M}_{e\bar e\to\nu_j^{}\bar\nu_j^{}\gamma}^{}\bigr|^2}  &=&
\frac{2e^2 g^4}{\kappa_-^{}\kappa_+^{}}\Bigl[
\bigl|{\cal G}_L^{jj}\bigr|^2\bigl(\hat u^2+\hat u^{\prime2}\bigr) +
\bigl|{\cal G}_R^{jj}\bigr|^2\bigl(\hat t^2+\hat t^{\prime2}\bigr) \Bigr] \hat s' ~, ~~~~
j \,=\, \mu,\tau ~,
\end{eqnarray}
\begin{eqnarray} \label{M2ee2nng''}
\overline{\bigl|{\cal M}_{e\bar e\to\nu_j^{}\bar\nu_l^{}\gamma}^{}\bigr|^2}  &=&
\frac{2 e^2 g^4}{\kappa_-^{}\kappa_+^{}}\Big[
\bigl|{\cal G}_L^{jl}\bigr|^2\bigl(\hat u^2+\hat u^{\prime2}\bigr) +
\bigl|{\cal G}_R^{jl}\bigr|\bigl(\hat t^2+\hat t^{\prime2}\bigr) \Bigr]\hat s' ~, ~~~~
j,l \,=\, e,\mu,\tau ~, ~~~~ j\,\neq\,l ~, ~~~~~~~
\end{eqnarray}
where
\begin{eqnarray} & \displaystyle
{\cal W}^{(\prime)} \,\,=\,\, \hat t^{(\prime)}-m_W^2+i\Gamma_W^{}m_W^{} ~, & \\ & \displaystyle
{\cal G}_{\sf C}^{ll'} \,\,=\,\,
\frac{\delta_{ll'}^{}\;\bar g_{\sf C}^{}}
{2c_{\rm w}^2\bigl(\hat s'-m_Z^2+i\Gamma_Z^{}m_Z^{}\bigr)} \,+\,
\frac{{\sf C}_{ll'}^{}}{g^2\bigl(\hat s'-m_X^2+i\Gamma_X^{}m_X^{}\bigr)} ~, \hspace{5ex}
{\sf C}^{} \,\,=\,\, L,R ~, &
\end{eqnarray}
with $\Gamma_W$ being the total width of $W$, etc.
In the numerical analysis, we use \,$\alpha=e^2/(4\pi)=1/128$,\,
\,$G_{\rm F}^{}=g^2/\bigl(32\,m_W^4\bigr){}^{1/2}=1.166\times10^{-5}{\rm\,GeV}^{-2}$,\, and
\,$\sin^2\theta_W^{}=0.23$.\,
With these parameters, we can reach most of the SM ranges listed in Table\,\ref{ee2nngdata}
to within~10\%.

\begin{table}[ht]
\caption{\baselineskip=13pt%
Measured and SM values of \,$e^+e^-\to\nu\bar\nu\gamma$\, cross section for various
$e^+e^-$ center-of-mass energies and cuts on
$E_\gamma{}$, \,$x=2E_\gamma^{}/\sqrt s$,\, $x_T^{}=x\sin\theta_\gamma^{}$,\, or
\,$E_{\gamma T}^{}=\sqrt s\,x_T^{}/2$\, and \,$\hat y=\cos\theta_\gamma^{}$.\,
The second (third) number in each $\sigma_{\rm exp}^{}$ entry is the statistical (systematic)
error. The $\sigma_{\rm SM}^{}$ entries are available from the experimental papers.
Most of these numbers were previously quoted in
Refs.\,\cite{Barranco:2007ej,Forero:2011zz,Hirsch:2002uv}.\label{ee2nngdata}} \footnotesize
\begin{tabular}{|c|c|ccl|}
\hline
& $\vphantom{\big|_|^|}\sqrt s$ (GeV) & $\sigma_{\rm exp}^{}$ (pb) &
$\sigma_{\rm SM}^{}$ (pb) & \hspace{5ex} $E_\gamma,\,x,\,x_T^{},\,\hat y$\, cuts \\
\hline\hline
ALEPH~\cite{Buskulic:1996hw}
& 130.0 & $9.6 \pm2.0 \pm0.3 $ &  $10.7\pm0.2$ &
\multirow{2}{*}{$\left.\begin{array}{c} \vspace{2ex} \\ \end{array}\right\}$
$E_\gamma\ge10$\,GeV, ~ $|\hat y|\le0.95$} \\
& 136.0 & $7.2 \pm1.7 \pm0.2 $ &  $9.1\pm0.2$ & \\ \hspace{8ex} \cite{Barate:1997ue}
& 161.0 & $5.3 \pm0.8 \pm0.2 $ &  $5.81\pm0.03$ &
\multirow{10}{*}{$\left.\begin{array}{c}\\ \\ \\ \\ \\ \\ \\ \\\vspace{2ex}\\\end{array}\right\}$
$x_T^{}\ge0.075$, ~ $|\hat y|\le0.95$} \\
& 172.0 & $4.7 \pm0.8 \pm0.2 $ &  $4.85\pm0.04$ & \\ \hspace{8ex} \cite{Barate:1998ci}
& 182.7 & $4.32\pm0.31\pm0.13$ &  $4.15\pm0.03$ & \\ \hspace{8ex} \cite{Heister:2002ut}
& 188.6 & $3.43\pm0.16\pm0.06$ &  $3.48\pm0.05$ & \\
& 191.6 & $3.47\pm0.39\pm0.06$ &  $3.23\pm0.05$ & \\
& 195.5 & $3.03\pm0.22\pm0.06$ &  $3.26\pm0.05$ & \\
& 199.5 & $3.23\pm0.21\pm0.06$ &  $3.12\pm0.05$ & \\
& 201.6 & $2.99\pm0.29\pm0.05$ &  $3.07\pm0.05$ & \\
& 205.0 & $2.84\pm0.21\pm0.05$ &  $2.93\pm0.05$ & \\
& 206.7 & $2.67\pm0.16\pm0.05$ &  $2.80\pm0.05$ & \\
\hline
\,DELPHI~\cite{Abreu:2000vk}\,
& 182.7 & $1.85\pm0.25\pm0.15$ &  $2.04\pm0.02$ &
\multirow{2}{*}{$\left.\begin{array}{c} \vspace{2ex} \\ \end{array}\right\}$
$x\ge0.06$, ~ $|\hat y|\le0.707$} \\
& 188.7 & $1.80\pm0.15\pm0.14$ &  $1.97\pm0.02$ & \\
& 182.7 & $2.33\pm0.31\pm0.19$ &  $2.08\pm0.02$ &
\multirow{2}{*}{$\left.\begin{array}{c} \vspace{2ex} \\ \end{array}\right\}$
$0.2\le x\le0.9$, ~ $0.848\le|\hat y|\le0.978$} \, \\
& 188.7 & $1.89\pm0.16\pm0.15$ &  $1.94\pm0.02$ & \\
& 182.7 & $1.27\pm0.25\pm0.11$ &  $1.50\pm0.02$ &
\multirow{5}{*}{$\left.\begin{array}{c} \\ \\ \\ \vspace{2ex} \\ \end{array}\right\}$
$0.3\le x\le0.9$, ~ $0.990\le|\hat y|\le0.998$ \,} \\
& 188.7 & $1.41\pm0.15\pm0.13$ &  $1.42\pm0.01$ & \\ \hspace{9ex} \cite{Abdallah:2003np}
& 187.1 & $1.37\pm0.14\pm0.11$ &  $1.44\pm0.01$ & \\
& 196.8 & $1.22\pm0.14\pm0.10$ &  $1.29\pm0.01$ & \\
& 205.4 & $1.12\pm0.11\pm0.09$ &  $1.18\pm0.01$ & \\
& 187.1 & $1.98\pm0.14\pm0.16$ &  $1.97\pm0.02$ &
\multirow{3}{*}{$\left.\begin{array}{c} \\ \vspace{2ex} \\ \end{array}\right\}$
$0.2\le x\le0.9$, ~ $0.848\le|\hat y|\le0.978$} \\
& 196.8 & $1.71\pm0.14\pm0.14$ &  $1.76\pm0.02$ & \\
& 205.4 & $1.71\pm0.12\pm0.14$ &  $1.57\pm0.02$ & \\
& 187.1 & $1.78\pm0.13\pm0.16$ &  $1.89\pm0.02$ &
\multirow{3}{*}{$\left.\begin{array}{c} \\ \vspace{2ex} \\ \end{array}\right\}$
$x\ge0.06$, ~ $|\hat y|\le0.707$} \\
& 196.8 & $1.41\pm0.13\pm0.13$ &  $1.75\pm0.02$ & \\
& 205.4 & $1.50\pm0.11\pm0.14$ &  $1.61\pm0.02$ & \\
\hline
 L3~\cite{Acciarri:1997dq}
& 161.3 & \, $6.75\pm0.91\pm0.18$ \, & $6.26\pm0.12$ &
\multirow{2}{*}{$\left.\begin{array}{c} \vspace{2ex} \\ \end{array}\right\}$
$E_{\gamma T}\ge6$\,GeV, ~ $|\hat y|\le0.97$} \\
& 172.3 & $6.12\pm0.89\pm0.14$ &  $5.61\pm0.10$ & \\ ~ ~ \cite{Acciarri:1998hb}
& 182.7 & $5.36\pm0.39\pm0.10$ &  $5.62\pm0.10$ &
\multirow{2}{*}{$\left.\begin{array}{c} \vspace{2ex} \\ \end{array}\right\}$
$E_\gamma\ge5$\,GeV, ~ $|\hat y|\le0.97$} \\ ~ ~ \cite{Acciarri:1999kp}
& 188.6 & $5.25\pm0.22\pm0.07$ &  $5.28\pm0.05$ & \\
\hline
 OPAL~\cite{Ackerstaff:1997ze}
& 130.3 & $10.0\pm2.3 \pm0.4 $ & $13.48\pm0.22$ &
\multirow{4}{*}{$\left.\begin{array}{c} \\ \\ \vspace{2ex} \\ \end{array}\right\}
\begin{array}{l} \mbox{$x_T^{}\ge0.05$, ~ $|\hat y|\le0.82$} \\ \hspace{10ex} \rm or \\
\mbox{$x_T^{}\ge0.1$, ~ $0.82\le|\hat y|\le0.966$} \end{array}$} \\ % 4% sys
& 136.2 & $16.3\pm2.8 \pm0.7 $ & $11.30\pm0.20$ & \\
& 161.3 & ~$5.3\pm0.8 \pm0.2 $ & ~$6.49\pm0.08$ & \\
& 172.1 & ~$5.5\pm0.8 \pm0.2 $ & ~$5.53\pm0.08$ & \\ ~ ~ ~ ~ ~ \cite{Abbiendi:1998yu}
& 130.0 & $11.6\pm2.5 \pm0.4 $ & $14.26\pm0.06$ &
\multirow{4}{*}{$\left.\begin{array}{c} \\ \\ \vspace{2ex} \\ \end{array}\right\}$
$x_T^{}\ge0.05$, ~ $|\!\cos\theta_\gamma|\le0.966$} \\
& 136.0 & $14.9\pm2.4 \pm0.5 $ & $11.95\pm0.07$ & \\
& 182.7 & $4.71\pm0.34\pm0.16$ & $ 4.98\pm0.02$ & \\ ~ ~ ~ ~ ~ \cite{Abbiendi:2000hh}
& 188.6 & $4.35\pm0.17\pm0.09$ & $ 4.66\pm0.03$ & \\
\hline\hline
\end{tabular}
\end{table}

In our framework, the $X$-mediated amplitude for the flavor-changing decay
\,$\ell_i\to\ell_j e^+e^-$\, proceeds from a one-loop diagram for \,$\ell_i\to\ell_j X^*$\,
involving internal $W$, $\nu_i^{}$, and $\nu_j^{}$ lines with $X$ attached to the neutrino lines
and eventually transforming~into $e^+e^-$.
Since the masses of the external and internal leptons are small relative to~$m_W^{}$, it is
a good approximation to retain only the lowest order terms in the small-mass expansion.
In that limit, we can employ the results of Ref.\,\cite{He:2009rz} to derive
\begin{eqnarray}
{\cal M}_{\ell_i\to\ell_j e^+e^-}^{} \,\,\sim\,\,
\frac{G_{\rm F}^{}\,m_\nu^2\,\ln\bigl(m_\nu^2/m_W^2\bigr)}{2\sqrt2\,\pi^2}\;
\frac{\bar\ell_j\gamma^\lambda P_L\ell_i\;\bar e\gamma_\lambda^{}(L_{ji}P_L+R_{ji}P_R)e}
{\hat s-m_X^2+i\Gamma_X^{}m_X^{}} ~,
\end{eqnarray}
after dropping a divergent term depending on the internal neutrino mass which in the complete
model would be canceled by other contributions, assuming the presence of a GIM-like mechanism in
the lepton sector of the model, and neglecting the final lepton masses.
This implies that we only have an order-of-magnitude estimate of the decay branching ratio,
given by
\begin{eqnarray} \label{l2lll}
{\cal B}\bigl(\ell_i\to\ell_j e^+e^-\bigr) \,\,\sim\,\,
\frac{8G_{\rm F}^2\,m_\nu^4\,\ln^2\bigl(m_\nu^2/m_W^2\bigr)\,\bigl(|L_{ji}|^2+|R_{ji}|^2\bigr)}
{3(4\pi)^7\,\Gamma_{\ell_i}\,m_{\ell_i}^3}\int_0^{m_{\ell_i}^2}d\hat s\;
\frac{\bigl(m_{\ell_i}^2-\hat s\bigr)^2\bigl(m_{\ell_i}^2+2\hat s\bigr)}
{\bigl(\hat s-m_X^2\mbox{$\bigr)^2$}+\Gamma_X^2m_X^2} ~. ~~~~
\end{eqnarray}
This is very suppressed for \,$m_\nu^{}<1$\,eV.\,

\newpage

%%%%%%%%%%%%%%%%%%%%%%%%%%%%%%%%%%%%%%%%%%%%%%%%%%

\end{document}